%% file: sliding_interface_29-11-2017.tex
\documentclass[11pt]{article}
\usepackage[top=1.0in,right=1.0in,left=1.0in,bottom=1.75in]{geometry}
\usepackage{amssymb,amsbsy}
\usepackage{amsmath}
\usepackage{amsthm}
\usepackage{graphics,graphicx}
\usepackage{color}
\usepackage[dvipsnames]{xcolor}
\usepackage{subfigure}

\usepackage[affil-it,auth-sc]{authblk}
\usepackage{stmaryrd}
\usepackage{booktabs}
\usepackage{array}

\usepackage{soul}

\newcolumntype{L}[1]{>{\raggedright\let\newline\\\arraybackslash\hspace{0pt}}m{#1}}
\newcolumntype{C}[1]{>{\centering\let\newline\\\arraybackslash\hspace{0pt}}m{#1}}
\newcolumntype{R}[1]{>{\raggedleft\let\newline\\\arraybackslash\hspace{0pt}}m{#1}}

\input{definitions}

\newcommand{\ljump}{\llbracket}
\newcommand{\rjump}{\rrbracket}
\newcommand{\jump}[1]{\ljump #1 \rjump}

\title{Bifurcation of elastic solids with sliding interfaces}

\date{}

\author[]{D. Bigoni${}^{\rm a}$, N. Bordignon${}^{\rm a}$, A. Piccolroaz${}^{\rm a}$, S. Stupkiewicz${}^{\rm b}$}

\affil[]{${}^{\rm a}$DICAM, University of Trento, Italy}
\affil[]{${}^{\rm b}$Institute of Fundamental Technological Research, Polish Academy of Sciences, Warsaw, Poland}

\begin{document}

\maketitle

\begin{abstract}
Lubricated sliding contact between soft solids is an interesting topic in biomechanics and for the design of small-scale engineering devices. As a model of this mechanical
set-up, two elastic nonlinear solids are considered jointed through a frictionless and bilateral surface, so that continuity of the normal component of the Cauchy traction holds 
across the surface, but the tangential component is null. 
Moreover, the displacement can develop only in a way that the bodies in contact do neither detach, nor overlap. 
Surprisingly, this finite strain problem has not been correctly formulated until now, so that this formulation is the objective of the present paper.
The incremental equations are shown to be non-trivial and different from previously (and erroneously) employed conditions. In particular, 
an exclusion condition for bifurcation is derived to show that previous formulations based on frictionless contact  or \lq spring-type' interfacial conditions are not able to predict bifurcations in tension, while experiments 
-- one of which, {\it ad hoc} designed, is reported -- show that these
bifurcations are a reality and become possible when the correct sliding interface model is used.
The presented results introduce a methodology for the  determination of bifurcations and instabilities occurring during lubricated sliding between soft bodies in contact. 
\end{abstract}

{\it Keywords: frictionless contact; large strains; nonlinear elasticity}

\section{Introduction}

\begin{figure}[!htb]
\centering
\includegraphics[width=12cm]{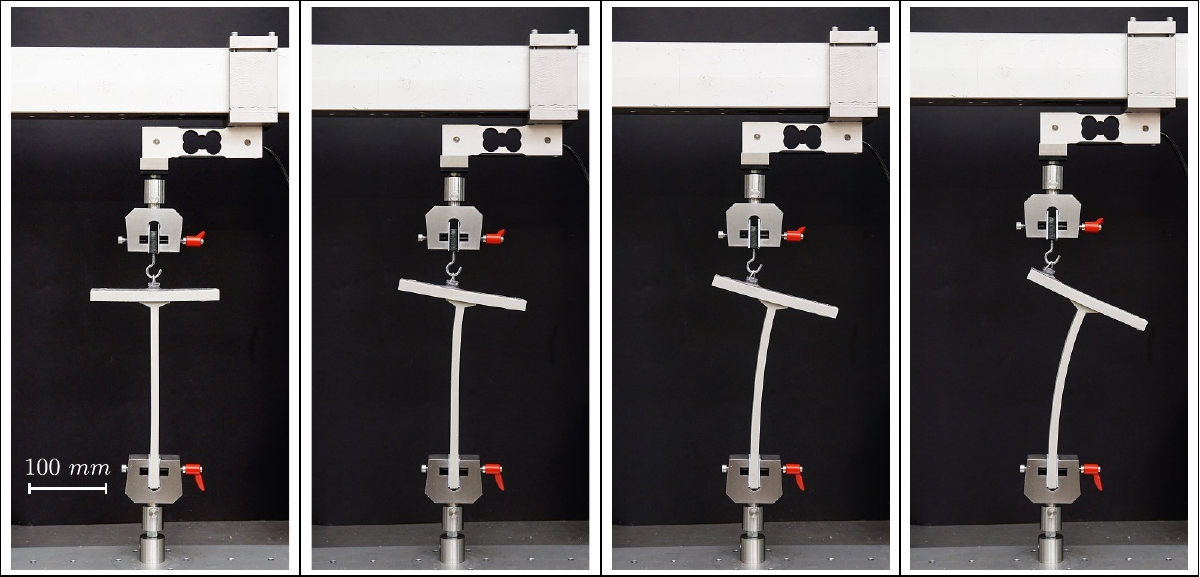}
\caption{\footnotesize{A sequence of photos showing a tensile bifurcation involving sliding contact between two soft solids. A silicon rubber suction cup is applied on a lubricant oil film to the upper part of a \lq T-shaped' silicon rubber (gray in the photo), clamped at the lower end. The suction cup is pulled vertically, so that the straight configuration of the \lq T' is a trivial equilibrium configuration (photo on the left) and 
a tensile bifurcation occurs when this element starts bending (second photo from the left) and the suction cup slips, as shown in the sequence of photos. Note that in this system 
rigid mechanical devices such as rollers or sliding sleeves are avoided. 
}}
\label{figFRAME}
\end{figure}

Lubricated sliding along an interface between two deformable bodies is typically characterized by very low friction and arises, for instance, 
in several 
biotribological systems (Dowson, 2012), such as the contact-lens/cornea (Dunn et al., 2013) and the articular cartilage (Ateshian, 2009) complexes, or 
in various engineering devices, such as windscreen wipers, aquaplaning tires,  and 
elastomeric seals (Stupkiewicz and Marciniszyn, 2009). 
These soft and slipping contacts are often characterized by large elastic or viscoelastic deformations so that it is not obvious how to formulate the Reynolds equation to adequately model the fluid flow between two contact surfaces that undergo large time-dependent deformations (Temizer and Stupkiewicz, 2016). 
Moreover, a distinctive feature of lubricated soft contacts is that they are capable of sustaining \emph{tensile contact tractions} during sliding, particularly in transient conditions, a phenomenon clearly visible when a suction cup is moved on a lubricated substrate. 
Indeed, as long as the pressure does not drop below the cavitation pressure, a soft contact can be loaded in tension, possibly imposing large deformations in a highly compliant solid.
As an example of this situation, 
the sequence of photos shown in Fig.~\ref{figFRAME} refers to an experiment (performed at the Instabilities Lab of the University of Trento) on tensile buckling involving a sliding contact between two soft solids. 
This system has been designed and realized to obtain a compliant sliding element, and thus to buckle in tension, without using rigid parts such as rollers or sliding sleeves. 
In particular, a \lq T-shaped' silicon rubber element is clamped at the lower end and connected at the upper flat end to a silicon rubber suction cup, which has been applied with a lubricant oil.
The system is pulled in tension and displays a tensile bifurcation in which the \lq T' bends while the suction cup slides along the upper flat end of the \lq T'. This bifurcation resembles that analyzed in (Zaccaria et al. 2011), but involves here soft solids.

A bilateral and frictionless sliding contact condition has been often employed to model the above-mentioned problems (for instance, in geophysics, Leroy and Triantafyllidis, 1996, or for sliding inclusions, Tsuchida et al., 1986, or roll-bonding of metal sheets, Steif, 1990), 
where two bodies in a current configuration share a common surface along which shear traction and normal separation/interpenetration must both vanish, but free sliding is permitted. 

Another model is based on a \lq spring-like' interface, in which the incremental nominal traction is related to the jump in the incremental displacement across the interface (see Suo et al.\ 1992; Bigoni et al.\ 1997). 
This model, in the limit of null tangential stiffness and null normal compliance should reduce to the sliding interface model. 
While these models are elementary within an infinitesimal theory, they become complex when the bodies in contact suffer large displacement/strain (and may evidence bifurcations, as in the case of the soft materials involved in the experimental set-up shown in Fig. \ref{figFRAME}). As a matter of fact, the freely sliding interface model has 
never been \emph{even formulated} so far and the \lq spring-like' model will be shown not to reduce to the freely sliding interface in the above-mentioned limit of vanishing tangential stiffness and normal compliance.

The correct formulation for a sliding interface, together with the derivation of incremental conditions, are the  focus of the present article: the former turns out to be non-trivial and the latter corrects  previously used conditions, which are shown to lead to incorrect conclusions. 
Moreover, a generalization of the Hill's exclusion condition for bifurcation (Hill, 1957; see Appendix \ref{appendixB}) to bodies containing interfaces, shows that the \lq spring-like' interface cannot explain 
bifurcations which can in fact be obtained with the correct formulation of the sliding contact and which exist in reality, as the above-mentioned experiment shows.

The availability of analytical solutions for incremental bifurcations of nonlinear elastic solids is crucial for many applications  involving soft materials 
(De Tommasi et al. 2010; Destrade and Merodio, 2011; deBotton et al. 2013; Ciarletta and Destrade, 2014; Steigmann and Ogden, 2014; Liang and Cai, 2015; Destrade et al. 2016; Riccobelli and Ciarletta, 2017), 
so that the importance of the model derived in this paper is that it allows to obtain solutions for bifurcations occurring in soft bodies in contact with a frictionless planar interface. Several of these solutions, which are important for applications, are here obtained, while 
other problems which do not admit an analytical solution are solved by employing the finite element method and a linear perturbation technique.
The 
obtained solutions show 
that sliding conditions strongly affect bifurcation loads and promote tensile bifurcations (such as that visible in the experiment reported in Fig. \ref{figFRAME}), which are shown to remain usually undetected by employing previously used, but incorrect, conditions.

\section{Sliding Interface Conditions}
\label{Sliding Interface Conditions}

\subsection{Problem formulation and kinematics of two bodies in frictionless contact}

\begin{figure}[!htb]
\centering
\includegraphics[width=130mm]{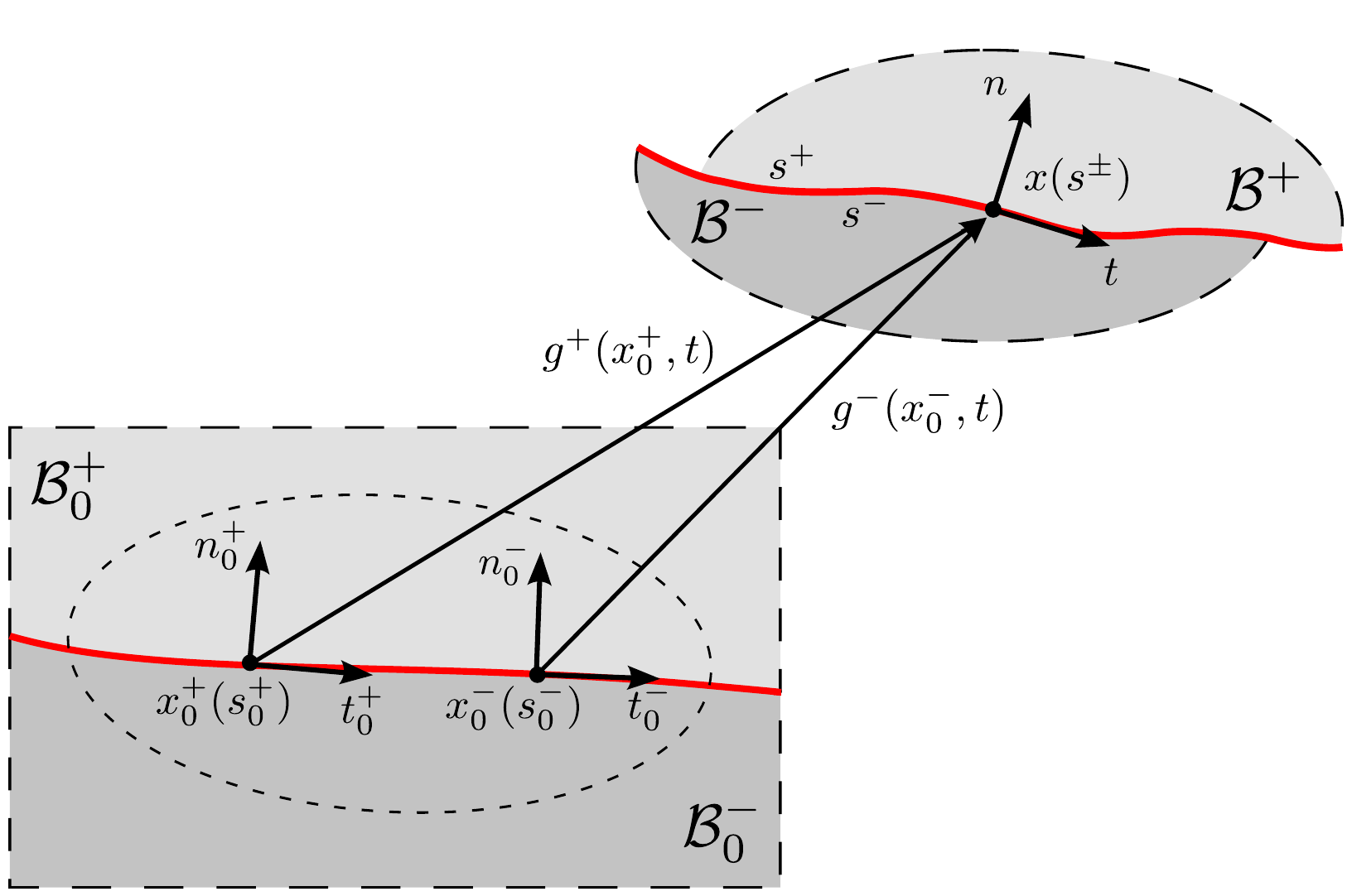}
\caption{\footnotesize Deformation of two nonlinear elastic bodies under plane strain conditions and jointed through a frictionless and bilateral interface. The interface constitutive law enforces a bilateral constraint on the  displacement (so that the two bodies can neither detach, nor interpenetrate, during deformation) and continuity of the 
Cauchy traction, but with the tangential component of the latter being null. A finite and unprescribed sliding of the two bodies can occur across the interface.}
\label{fig01}
\end{figure}

Two nonlinear elastic bodies (denoted by \lq $+$' and \lq $-$') are considered in \emph{plane-strain} conditions, jointed through a bilateral frictionless interface, Fig.~\ref{fig01}. 
Points in the reference configurations $\mathcal B_{0}^+$ and $\mathcal B_{0}^-$ are mapped to the current configurations $\mathcal B^+$ and $\mathcal B^-$ via the deformations
$\bg^{\pm} : \mathcal B_{0}^{\pm}  \rightarrow \mathcal B^{\pm} $, so that
\begin{equation}
\bx^{+} = \bg^{+} ( \bx_{0}^{+} , t ), 
~~~~
\bx^{-} = \bg^{-} ( \bx_{0}^{-} , t ),
\end{equation}
where $t$ denotes the time, the subscript \lq $0$' is used to highlight the referential description. Therefore,  the displacement vector $\bu$ is related to the deformation through
\beq
\bu^\pm = \bg^{\pm} ( \bx_{0}^{\pm} , t )-\bx_0^\pm
\eeq
where \lq $\pm$' denotes that the equation holds for both quantities \lq $+$' and \lq $-$'. 

The interface has the form of a regular surface $\Sigma$ in the current configuration and is the image of another regular surface $\Sigma_0$ in the reference configuration, where it admits the arc-length parameterization
\begin{equation}
\label{param}
\bx_{0}^{+} = \bx_{0} ( s_{0}^{+} ) , 
\end{equation}
so that, since the parameter $s_{0}^{-}$ can be expressed as function of $s_{0}^{+}$ and time, the following expression can be derived
\beq
\label{param2}
\bx_{0}^{-} = \bx_{0} ( s_{0}^{-} ) = \bx_{0} ( s_{0}^{-} ( s_{0}^{+} , t ) ).
\eeq

The unit tangent vectors to the surface in the reference configuration, $\Sigma_0$, can be expressed as 
\begin{equation}
\label{tgzeropm}
\bt_{0}^{+} = \frac{\partial {\bx_{0}^{+}}}{\partial {s_{0}^{+}}} \frac{1}{|\frac{\partial {\bx_{0}^{+}}}{\partial {s_{0}^{+}}}|} , \qquad 
\bt_{0}^{-} = \frac{\partial {\bx_{0}^{-}}}{\partial {s_{0}^{-}}} \frac{1}{ | \frac{\partial {\bx_{0}^{-}}}{\partial {s_{0}^{-}}} | } .
\end{equation}

Note that a point $\bx$ on the interface $\Sigma$ in the current configuration is the image of two different points $\bx^+_0$ and $\bx^-_0$ on $\Sigma_0$. This condition, {\it representing the fact that the two bodies in contact can neither detach nor interpenetrate}, can be expressed as $ \bx = \bx^{+} = \bx^{-} $ so that
\begin{equation}
\label{gpgm}
\bg^{+} ( \bx_{0}^{+} ( s_{0}^{+} ) , t ) = \bg^{-} ( \bx_{0}^{-} ( s_{0}^{-} ( s_{0}^{+} , t ) ) , t ) .
\end{equation}

The above condition defines the implicit dependence of $s_{0}^{-}$ on $s_{0}^{+}$ (and time) that has already been exploited in Eq.\ (\ref{param2}). 
Introducing the deformation gradient
\beq
\bF^\pm = \frac{\partial \bg^\pm}{\partial \bx^\pm_0} ,
\eeq
taking the derivative of Eq.\ (\ref{gpgm}) with respect to $s_{0}^{+}$ and applying the chain rule of differentiation yields 
\begin{equation}
\bF^{+} \frac{ \partial \bx_{0}^{+} }{ \partial s_{0}^{+} } = \bF^{-} \frac{ \partial \bx_{0}^{-} }{ \partial s_{0}^{-} } \frac{ \partial s_{0}^{-} }{ \partial s_{0}^{+} }, 
\end{equation}
finally leading to the definition of the tangent vector $\bt$ in the spatial configuration on $\Sigma$ at $\bx$
\begin{equation}
\bt = \frac{\bF^{+} \bt_{0}^{+}}{| \bF^{+} \bt_{0}^{+} |} = \frac{\bF^{-} \bt_{0}^{-}}{| \bF^{-} \bt_{0}^{-} |}. 
\label{ttt}
\end{equation}

The unit normal at $\bx$ on $\Sigma$ can be obtained through the Nanson's rule of area transformation
\begin{equation}
\lb{nan}
\bn = \frac{ A_{0}^+} {A^+} J^+ (\bF^+)^{-T} \bn^+_{0} = \frac{ A_{0}^- } {A^-} J^- (\bF^-)^{-T} \bn^-_{0},
\end{equation}
so that 
\begin{equation}
\bn = \frac{(\bF^{+})^{-T} \bn_{0}^{+}}{| (\bF^{+})^{-T} \bn_{0}^{+} |} = \frac{(\bF^{-})^{-T} \bn_{0}^{-}}{| (\bF^{-})^{-T} \bn_{0}^{-} |} .
\label{nnn}
\end{equation}

Note that while $\bn_0^+$ and $\bn_0^-$, as well as $\bt^+_0$ and $\bt^-_0$, are different, there is only one $\bn$ and one $\bt$.

\subsection{Tractions along the sliding interface} 

The interface is assumed to maintain a frictionless sliding contact, so that the normal component of the Cauchy traction has to be continuous and the tangential component null. These conditions can be written as follows
\beq
\lb{lapeppina}
\bn \scalp \jump{ \bT } \bn = 0,  \qquad
\bt \scalp \bT^{+} \bn = \bt \scalp \bT^{-} \bn  = 0,
\eeq
where $\bT$ is the Cauchy stress and $\jump{ \aleph } = \aleph^+-\aleph^-$ is the jump operator of the quantity $\aleph$ across $\Sigma$.
On introduction of the first Piola--Kirchhoff stress $\bS = J \bT\bF^{-T}$ (where $J=\det \bF$) and using the Nanson's rule (\ref{nan}) yields 
\begin{equation}
\bT^\pm \bn = \frac{\bS^\pm \bn_{0}^\pm}{\iota^\pm}, 
\label{relazTS}
\end{equation}
where $\iota^\pm = A^\pm/A_0^\pm$ is the ratio between the spatial and referential area elements, so that Eqs.\ (\ref{lapeppina}) can be transformed to
\beq
\lb{failcaffe}
\bn \scalp \left( \frac{\bS^+ \bn_{0}^+}{\iota^+} - \frac{\bS^- \bn_{0}^-}{\iota^-} \right) = 0,
\qquad
\bt \scalp  \frac{\bS^+ \bn_{0}^+}{\iota^+} = \bt\scalp \frac{\bS^- \bn_{0}^-}{\iota^-} = 0.
\eeq

\subsection{Motion of two solids in frictionless contact} 

Before deriving the relations pertaining to the interface, the following relations are introduced which are standard for continua and still hold for points at the left and right limit of $\Sigma$: 

\begin{itemize}

\item The material time derivative, denoted by a superimposed dot, of the tangent and normal unit vectors to the surface $\Sigma$ at $\bx$ is
\begin{equation}
\dot {\bt}^{\pm} = (\bI - \bt \otimes \bt) \bL^{\pm} \bt, 
\label{tpunto}
\end{equation}
\begin{equation}
\dot {\bn}^{\pm} = - (\bI - \bn \otimes \bn) (\bL^\pm)^{T} \bn
\label{npunto}
\end{equation}
where $ \bI $ is the identity tensor, $ \bL^{\pm} $ is the gradient of the spatial description of velocity for the \lq $+$' and \lq $-$' parts of the body
\begin{equation}
\bL^{\pm} (\bx^{\pm},t) = \grad \bv^{\pm} ,
\end{equation} 
and $\bv$ is the spatial description of the velocity 
\begin{equation}
\bv^{\pm} ( \bx^{\pm},t) = \dot {\bx}^{\pm} ( \bx_0^\pm ( \bx^{\pm} ,t ) ,t ) ,
\end{equation} 
where $\bx_0^\pm=\bx_0^\pm(\bx^\pm,t)$ denotes the inverse of $\bx^\pm=\bg^\pm(\bx_0^\pm,t)$.

\item The ratio between the deformed and the undeformed area elements can be obtained from the Nanson's rule, Eq.~(\ref{nan}), as 
\begin{equation}
\iota^{\pm} = J^{\pm} \big| (\bF^\pm)^{-T} \bn_{0}^{\pm} \big| , 
\end{equation} 
from which its material time derivative can be obtained in the form 
\begin{equation}
\dot{\iota}^{\pm} = J^{\pm} ( \tr \bL^{\pm}  - \bn \scalp \bL^{\pm} \bn ) \, \big| (\bF^\pm)^{-T} \bn_{0}^{\pm} \big|
 {} = \iota^{\pm} (\bI - \bn \otimes \bn) \scalp \bL^{\pm}  ,
\end{equation} 
as well as the following material time derivative 
\begin{equation}
\left( \frac{1}{\iota^{\pm}} \right)^{\scalp} = \frac{- \tr \bL^{\pm} + \bn \scalp \bL^{\pm} \bn }{J^{\pm} \vert (\bF^\pm)^{-T} \bn_{0}^{\pm} \vert } 
 {} = -\frac{1}{\iota^{\pm}} (\bI - \bn \otimes \bn) \scalp \bL^{\pm}  .
\end{equation}

\end{itemize}

A point on the sliding interface $\Sigma$ has to be understood as the \lq superposition' of the two points, one belonging to the body $\mathcal B^+$ and the other to the body $\mathcal B^-$, so that $\bx^+=\bx^-$ along $\Sigma$. Taking the
time derivative of the equation $\bx^+=\bx^-$ at fixed $s_{0}^{+}$, the velocities of the two points $\bx^+$ and $\bx^-$ can be related to each other through
\begin{equation}
\dot{\bx}^{+} = \dot{\bx}^{-} + \bF^{-} \frac{ \partial \bx_{0}^{-} }{ \partial s_{0}^{-} } \dot {s}_{0}^{-} .
\label{condi1}
\end{equation}
The time derivative at fixed $s_{0}^{+}$ is in fact the material time derivative for the \lq $+$' part of the body, while it involves an additional term related to the variation of $s_{0}^{-}$ for the \lq $-$' part of the body.

Equations (\ref{tgzeropm}) and (\ref{ttt})  show that $\bF^- \partial \bx_{0}^{-} /\partial s_{0}^{-}$ is parallel to the tangent unit vector $\bt$, so that the scalar product of the unit normal $\bn$ with both sides of Eq.~(\ref{condi1}) yields the continuity condition across the interface $\Sigma$ for the normal component of the velocity 
\begin{equation}
\label{xdot}
\jump{ \dot{\bx}}\scalp\bn = 0 ,
\end{equation}
while the scalar product with the unit tangent $\bt$ yields $\dot {s}_{0}^{-}$, thus
\begin{equation}
\dot{s}_{0}^{-} = 
\frac{( \dot {\bx}^{+} - \dot {\bx}^{-} ) \scalp \bt}{\vert \bF^{-} \frac{\partial \bx_{0}^{-}}{\partial s_{0}^{-}} \vert} .
\end{equation}

The time derivative of Eqs.~\eqref{ttt} and \eqref{nnn} at fixed $s_{0}^{+}$ provides
\begin{equation}
\dot{\bt}^{+} = \dot{\bt}^{-} + \frac{\partial \bt}{\partial s_{0}^{-}} \dot{s}_{0}^{-},
~~~~~~
\dot{\bn}^{+} = \dot{\bn}^{-} + \frac{\partial \bn}{\partial s_{0}^{-}} \dot{s}_{0}^{-},
\eeq
which using Eqs.~(\ref{tpunto}) and (\ref{npunto}) lead to 
\beq
\frac{\partial \bt}{\partial s_{0}^{-}} \dot{s}_{0}^{-} = (\bI - \bt \otimes \bt ) \jump{\bL} \bt,
~~~~~~
\frac{\partial \bn}{\partial s_{0}^{-}} \dot{s}_{0}^{-} = - (\bI - \bn \otimes \bn ) \jump{\bL^{T}} \bn.
\label{nspunto}
\end{equation}

The scalar product of Eqs.\ (\ref{nspunto}) with $ \bt $ and $ \bn $ yields 
\begin{equation}
\bt \scalp \frac{\partial \bt}{\partial s_{0}^{-}} \dot{s}_{0}^{-} = 0 , \qquad 
\bn \scalp \frac{\partial \bt}{\partial s_{0}^{-}} \dot{s}_{0}^{-} = \jump{ L_{nt}} ,
\label{transp1}
\end{equation}
and 
\begin{equation}
\bn \scalp \frac{\partial \bn}{\partial s_{0}^{-}} \dot{s}_{0}^{-} = 0 , \qquad 
\bt \scalp \frac{\partial \bn}{\partial s_{0}^{-}} \dot{s}_{0}^{-} = - \jump{ L_{nt}} .
\label{transp2}
\end{equation}

The time derivative of Eq.~(\ref{failcaffe})$_1$ at fixed $s_{0}^{+}$ allows to obtain
\begin{multline}
\label{pippona}
\bn \scalp \frac{ \dot{\bS}^{+} \bn_{0}^{+}}{\iota^{+}} 
- \bn \scalp \frac{ \dot{\bS}^{-} \bn_{0}^{-}}{\iota^{-}} 
- \dot{s}_{0}^{-} 
\biggl(
\bn \scalp \frac{ \partial \bS^{-} }{ \partial s_{0}^{-}} \frac{ \bn_{0}^{-} }{\iota^{-}}
+\bn \scalp \bS^{-} \bn_{0}^{-} \frac{ \partial \left( \frac{1}{\iota^{-}}\right)}{ \partial s_{0}^{-}} \\
+ \bn \scalp \frac{ \bS^{-}}{\iota^{-}} \frac{\partial \bn_{0}^{-}}{\partial s_{0}^{-} } 
\biggl)
=
\bn \scalp \bT \bn \, \jump{L_{tt}} ,
\end{multline}
while the time derivative of Eq.~(\ref{failcaffe})$_2$ at fixed $s_{0}^{+}$ leads to
\begin{equation}
\bt  \scalp \dot{\bS}^{+} \bn_{0}^{+} = - \dot{\bt}^+  \scalp \bS^{+} \bn_{0}^{+} 
\end{equation}
and
\begin{equation}
\bt \scalp \dot{\bS}^{-} \bn_{0}^{-} 
= -\dot{\bt}^{-} \scalp \bS^{-} \bn_{0}^{-} - \dot{s}_{0}^{-} \frac{\partial \bt^{-}}{\partial s_{0}^{-}} \scalp \bS^{-} \bn_{0}^{-} -  \dot{s}_{0}^{-} \bt^{-} \scalp \frac{\partial \bS^{-}}{\partial s_{0}^{-}}\bn_{0}^{-} - \dot{s}_{0}^{-}  \bt^{-} \scalp \bS^{-} \frac{\partial \bn_{0}^{-}}{\partial s_{0}^{-}} ,
\end{equation}
so that, using Eqs.~\eqref{transp1}, \eqref{transp2}, and \eqref{tpunto}, the following expressions are derived
\begin{equation}
\label{pippa1}
\bt  \scalp \dot{\bS}^{+} \bn_{0}^{+}  = - L_{nt}^{+} \bn \scalp \bS^{+} \bn_{0}^{+},
\end{equation}
and
\begin{equation}
\label{pippa2}
\bt \cdot \dot{\bS}^{-} \bn_{0}^{-} = - L_{nt}^{+} \bn \cdot \bS^{-} \bn_{0}^{-} -  \dot{s}_{0}^{-} \bt^{-} \cdot \frac{\partial \bS^{-}}{\partial s_{0}^{-}}\bn_{0}^{-} -  \dot{s}_{0}^{-} \bt^{-} \cdot \bS^{-} \frac{\partial \bn_{0}^{-}}{\partial s_{0}^{-}} .
\end{equation}

\section{Planar Sliding Interface Conditions}
\label{Planar Sliding Interface Conditions}

The general interface conditions derived above are now simplified for the special case of a planar sliding interface that is assumed to satisfy the following conditions:
\begin{itemize}
\item the interface is planar both in the reference and in the current configurations (but can {\it incrementally} assume any curvature), so that:
\begin{equation}
\bn = \bn_{0}^{+} = \bn_{0}^{-}  , \qquad \bt = \bt_{0}^{+} = \bt_{0}^{-} , \qquad
\frac{\partial \bn_{0}^{-}}{\partial s_{0}^{-}} = \b0 ;
\end{equation}

\item the Cauchy traction components are uniform at the interface and satisfy:
\beq
T^+_{nn} = T^-_{nn} , \qquad T^+_{nt} = T^-_{nt} = 0 ;
\eeq

\item a relative Lagrangian description is assumed in which the current configuration is assumed as reference (so that $\bF^+=\bF^-=\Id$ and $\iota^+=\iota^-=1$ and $\bS^\pm = \bT^\pm$).

\end{itemize}
It follows from the above assumptions that 
\begin{equation}
\frac{ \partial \left( \frac{1}{\iota^{-}}\right)}{ \partial s_{0}^{-}} = 0 , \qquad 
\frac{\partial \bS^{-}}{\partial s_{0}^{-}} = \b0 .
\end{equation}

Now, introducing a reference system $x_1$--$x_2$ aligned parallel respectively to the unit tangent $\bt$ and normal $\bn$ to the interface, the equations governing the \emph{rate} problem across the above-introduced planar interface are the following:
\begin{itemize}
\item continuity of normal incremental displacements, from Eq.~(\ref{xdot}),
\begin{equation}
\lb{ball}
\dot{x}_{n}^{+} \left( x_{1},0 \right) = \dot{x}_{n}^{-} \left( x_{1},0 \right);
\end{equation}
\item continuity of incremental nominal shearing accross the interface, from Eqs.~(\ref{pippa1}) and (\ref{pippa2}),
\begin{equation}
\lb{ballred}
\dot{S}_{tn}^{+} \left( x_{1},0 \right) = \dot{S}_{tn}^{-} \left( x_{1},0 \right) ;    
\end{equation}
\item dependence of the incremental nominal shearing on the Cauchy stress component orthogonal to the interface $T_{nn}$ and incremental displacement gradient mixed component $L_{nt}$, 
from Eq.~(\ref{pippa1}),
\begin{equation}
\lb{ballgreen}
\dot{S}_{tn}^{+} \left( x_{1},0 \right) = - \alpha T_{nn} L_{nt} \left( x_{1},0 \right) ,
\end{equation}
where $\alpha=1$;
\item dependence of the jump in the incremental nominal stress orthogonal to the interface on the Cauchy normal component $T_{nn}$ and the jump in the tangential component of the incremental displacement gradient $L_{tt}$, from Eq.~(\ref{pippona}),
\begin{equation}
\lb{ballblack}
\dot{S}_{nn}^{+} \left( x_{1},0 \right) - \dot{S}_{nn}^{-} \left( x_{1},0 \right) = \alpha  T_{nn} \jump{L_{tt} \left( x_{1},0 \right)} .
\end{equation}
where, again, $\alpha=1$.
\end{itemize}

The parameter $\alpha$ has been introduced in the above equations to highlight the difference with respect to the incorrect conditions 
sometimes assumed at the interface (for instance by Steif, 1990)
\beq
\dot{S}^{\pm}_{tn}(x_1,0) = 0 , \qquad \dot{S}^+_{nn} = \dot{S}^-_{nn}, 
\eeq
which correspond to $\alpha=0$. Note that the only possibility to obtain a coincidence between the correct $\alpha=1$ and the incorrect $\alpha=0$ conditions is when the stress normal to the interface vanishes, namely, when $T_{nn}=0$.

The \lq spring-type' interfacial conditions used by Suo et al. (1992), Bigoni et al. (1997) and Bigoni and Gei (2001) do not reduce (except when $T_{nn}=0$) to the correct frictionless sliding conditions (\ref{ballgreen}) and (\ref{ballblack}), in the limit when the stiffness tangential to the interface tends to zero and the normal 
stiffness to infinity. In this limit case, the \lq spring-type' conditions reduce to the incorrect equations obtained with $\alpha=0$, so that they cannot properly describe slip without friction, unless when $T_{nn}=0$. Note that the stress orthogonal to the interface, $T_{nn}$ has been always assumed to be null by Bigoni et al. (1997) and Bigoni and Gei (2001); all bifurcation analyses reported in these papers are therefore different from those considered in the present paper, where the transverse stress is never null.

\subsection{Plane strain bifurcation problems involving a planar interface}

In the following, a series of incremental bifurcation problems are solved, involving two elastic nonlinear solids in contact through a sliding interface aligned parallel to the $x_1$--axis.
This problem set-up is similar to various situations analyzed in the literature 
(Dowaikh and Ogden, 1991; Cristescu et al. 2004; Ottenio et al., 2007; Fu and Cai, 2015; Fu and Ciarletta, 2015), with the variant that now the interfacial conditions are different.
It is important to highlight that the two solids in contact may be characterized by different constitutive assumptions and may be subject to a 
different state of prestress in the $x_1$--direction. In fact, the possibility that the two bodies may freely slide across the interface allows to 
relax the usual compatibility restrictions. 

The incremental constitutive equations are characterized by the following parameters (Bigoni, 2012, Chapter 6.2)
\begin{equation}
\xi = \frac{\mu^{*}}{\mu} , \qquad \eta = \frac{T_{tt} + T_{nn}}{2 \mu} , \qquad k = \frac{T_{tt} - T_{nn}}{2 \mu} ,
\label{nina}
\end{equation}
so that 
\begin{equation}
\begin{array}{ll}
\dot{S}_{11} = \mu (2\xi -k-\eta) L_{11} + \dot{p}, &~~~ \dot{S}_{22} = \mu (2\xi +k-\eta) L_{22} + \dot{p}, \\ [5 mm]
\dot{S}_{21} = \mu [(1+k) L_{21} + (1-\eta) L_{12}], &~~~ \dot{S}_{12} = \mu [(1-\eta) L_{21} + (1-k) L_{12}], 
\end{array}
\end{equation}
where $\dot{p}$ plays the role of a 
Lagrange multiplier, because the body is assumed incompressible, $L_{kk}=0$. 
For the sake of simplicity, a neo-Hookean material behaviour is assumed, $ \xi = 1 $, so that the material always lies in the elliptic imaginary (EI) regime and 
\begin{equation}
-1< k <1 , \qquad \Lambda = \sqrt{4\xi^2-4\xi+k^2}= | k | ,
\end{equation}
together with additional definitions to be used later,
\begin{equation}
\beta_{1} = \sqrt{ \frac{1 + |k| }{1 - |k|}}\quad \beta_{2} = \sqrt{ \frac{1 - |k| }{1 - |k|}},
\quad \quad
\Omega_{1} = i \beta_1 , \quad \Omega_{2} = i \beta_2,
\quad  
\Omega_{3} = - i \beta_1, \quad \Omega_{4} = - i \beta_2 .
\end{equation}

\subsubsection{Two elastic prestressed half-spaces in contact through a planar sliding interface} 
\label{sec:half-spaces}

\begin{figure}[!htb]
\centering
\includegraphics[width=16.5cm]{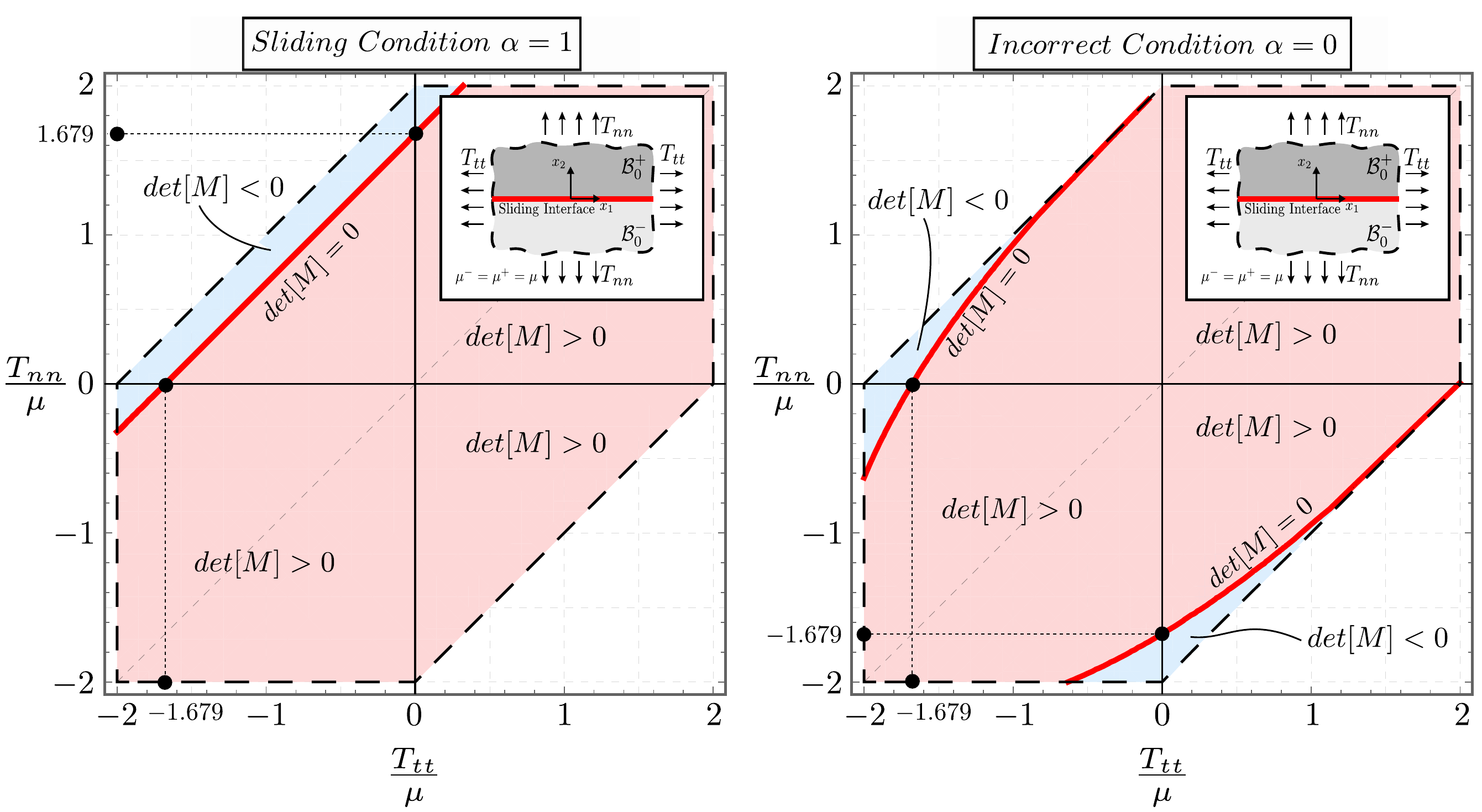}
\caption{\footnotesize{Interfacial bifurcation of two elastic incompressible half-spaces (made up of the same neo-Hookean material, subject to the same prestress) in contact through a planar sliding interface in the $ T_{nn} {-} T_{tt} $ plane for a sliding interface $\alpha=1$ (left). 
The incorrect condition $\alpha=0$ is also included for comparison (right). 
The points corresponding to bifurcation are represented by red lines (at the boundary between the red and blue zones), while the dashed lines correspond to failure of ellipticity. 
Note that with $\alpha=1$ bifurcation in pure tension occurs (i.e. with $T_{tt}=0$), which is excluded for $\alpha=0$. Therefore, the (correct) sliding interface condition explains tensile bifurcation. Note also that in this case bifurcations for both negative stresses $T_{nn}$ and $T_{tt}$ do not occur
(except in the domain of slightly negative $T_{nn}$).
}}
\label{PIANOcont}
\end{figure}

Two elastic half-spaces are now considered in contact through a sliding interface, planar in the current configuration, which is assumed as 
reference configuration, see the inset in Fig.~\ref{PIANOcont}. 

The upper (the lower) half-space $ x_{2} > 0 $ $ (x_{2} < 0) $ is denoted with \lq +' (with \lq$-$') and the incremental conditions at the interface 
are given by Eqs.~(\ref{ball})--(\ref{ballblack}), 
plus the condition of exponential decay of the solution in the limits $ x_{2} \rightarrow \pm \infty $. 
For simplicity the two half spaces are modelled with the same material and subject to the same prestress, so that bifurcations are possible only due to the presence of the interface. 

Employing the representation 
\begin{equation}
v_1^{\pm} = \widetilde{v}_1^{\pm} (x_2) f ( c_1,x_1) , 
~~~~
v_2^{\pm} = \widetilde{v}_2^{\pm} (x_2) f^{'} ( c_1,x_1),
\end{equation}
\begin{equation}
f ( c_1,x_1) = \exp (i c_1 x_1) , 
~~~~
f^{'} ( c_1,x_1) = i f ( c_1,x_1),
\end{equation}
\begin{equation}
\widetilde{v}_1^{\pm} (x_2) = -b_1^{\pm} \Omega_1^{\pm} e^{i c_1 \Omega_1^{\pm} x_2} -b_2^{\pm} \Omega_2^{\pm} e^{i c_1 \Omega_2^{\pm} x_2} -b_3^{\pm} \Omega_3^{\pm} e^{i c_1 \Omega_3^{\pm} x_2} -b_4^{\pm} \Omega_4^{\pm} e^{i c_1 \Omega_4^{\pm} x_2}  ,
\end{equation}
\begin{equation}
\widetilde{v}_2^{\pm} (x_2) = - i \left [ b_1^{\pm} e^{i c_1 \Omega_1^{\pm} x_2} + b_2^{\pm} e^{i c_1 x_2} + b_3^{\pm} e^{i c_1 \Omega_3^{\pm} x_2} + b_4^{\pm} e^{i c_1 \Omega_4^{\pm} x_2}  \right]
\end{equation}
for the incremental displacements (Bigoni, 2012), where $c_1$ is the wavenumber of the bifurcated mode, the decaying condition implies
\begin{equation}
b_{1}^{-} = b_{2}^{-} = b_{3}^{+} = b_{4}^{+} = 0 ,
\end{equation}
so that the eigenvalue problem governing incremental bifurcations can be written as
\begin{equation}
\begin{bmatrix} 
\bM \end{bmatrix} 
\begin{bmatrix} 
b_{1}^{+} \\ 
b_{2}^{+} \\
b_{3}^{-} \\ 
b_{4}^{-} \\
\end{bmatrix} = 0 ,
\label{sitemall}
\end{equation}
where the matrix $[\bM]$ is given by
\begin{equation}
\scalebox{0.75}{$
\begin{bmatrix} 
1 &  1 & -1  & -1 \\
2 - \eta + \Lambda  & 2 - \eta - \Lambda & - 2 + \eta - \Lambda & - 2 + \eta + \Lambda \\
2 - \eta + \Lambda + \frac{T_{nn}}{\mu} \alpha & 2 - \eta - \Lambda + \frac{T_{nn}}{\mu} \alpha & 0 & 0 \\
\left( 2 - \eta - \Lambda + \frac{T_{nn}}{\mu} \alpha \right) \sqrt{ \frac{1 + \Lambda }{1 - k}}     & 
\left( 2 - \eta + \Lambda + \frac{T_{nn}}{\mu} \alpha \right)  \sqrt{ \frac{1 - \Lambda }{1 - k}}    & 
\left( 2 - \eta - \Lambda + \frac{T_{nn}}{\mu} \alpha \right) \sqrt{ \frac{1 + \Lambda }{1 - k}}    & 
\left( 2 - \eta + \Lambda + \frac{T_{nn}}{\mu} \alpha \right) \sqrt{ \frac{1 - \Lambda }{1 - k}}    \\
\end{bmatrix}$
}.
\end{equation}

Non-trivial solutions of the system \eqref{sitemall} are obtained when $\det\bM=0$, to be solved 
for the bifurcation stress. Note that matrix $ \bM  $ does not contain the wavenumber of the bifurcated mode, so that the critical load for bifurcation is independent of the wavelength of the 
bifurcation mode (even if the sliding interface is present).

The resulting bifurcation condition for a sliding interface $ ( \alpha = 1 ) $ can be written as
\begin{equation} 
\sqrt{1-\Lambda} \left( \frac{T_{nn}}{\mu} + 2 -\eta + \Lambda \right)^{2} - \sqrt{ 1 +\Lambda} \left( \frac{T_{nn}}{\mu} + 2 -\eta - \Lambda \right)^{2}
=0. 
\label{eqcentrale}
\end{equation}
If, instead of the correct interface conditions, $\alpha=1$, one assumes the incorrect condition $\alpha = 0 $, bifurcation corresponds to 
\begin{equation}
\sqrt{1-\Lambda} \left( 2 -\eta + \Lambda \right)^{2} - \sqrt{ 1 +\Lambda} \left( 2 -\eta - \Lambda \right)^{2}
=0. 
\label{crackazzo}
\end{equation}
Using Eqs.~\eqref{nina} and for given values of longitudinal $ T_{tt} $ and transverse $T_{nn}$ prestresses, Eqs.~\eqref{eqcentrale} and (\ref{crackazzo}) (which hold for a generic incompressible material, subject to generic prestress conditions) can be solved. Results are reported in Fig.~\ref{PIANOcont} for a neo-Hookean material, $\xi=1$, assuming both the correct condition $\alpha=1$ (on the left) and the incorrect one $\alpha=0$ (on the right).
The red and blue zones identify in the figure the prestress combinations for which $\det\bM$ assumes positive and negative values, respectively, so that the boundary 
between these zones (marked with red lines) corresponds to bifurcation.
The dashed lines represent failure of ellipticity, so that points situated beyond this line do not represent states attainable through a smooth deformation path 
(because ellipticity loss corresponds to the emergence of discontinuous solution).

Note that in the case of null prestress normal to the interface, $T_{nn}=0$, an interfacial bifurcation occurs for $T_{tt}/\mu \approx -1.679$, the same value which gives the surface instability of a half space, which is unaffected by the condition $\alpha=1$ or $\alpha=0$. This is the only situation in which the two conditions provide the same bifurcation stress.

An interesting case occurs when only a tensile prestress orthogonal to the interface $T_{nn}$ is applied (and the transverse prestress is null, $T_{tt}=0$), where a {\it tensile bifurcation} occurs for $T_{nn}/\mu \approx$ 1.679, which is absent when the incorrect condition $\alpha=0$ is used or also if the modelling would involve a perfectly bonded interface (in which case all bifurcations are excluded within the limits of ellipticity). This simple example reveals the importance of a correct definition of the interfacial conditions. 

A comparison between the correct $\alpha=1$ and incorrect $\alpha=0$ conditions reveals a completely different bifurcation behaviour. In fact, for positive $T_{nn}$ bifurcation is possible in the correct case for negative, null and slightly positive $T_{tt}$. These bifurcations do not occur in the incorrect situation. Moreover in the latter situation there is a zone of bifurcation occurring for negative $T_{nn}$ which is excluded in the correct case. 
As an example, in the special, but interesting, case of uniaxial compression ($T_{nn}<0$ with $T_{tt}=0$), there is no bifurcation in the correct case $\alpha=1$, while bifurcation occurs in the other case.

To better elucidate this situation, an exclusion condition of the Hill (1957) type is derived in Appendix \ref{appendixB}. 
For $\alpha=0$, this condition becomes completely insensible to the presence of the sliding interface (and reduces to the Hill's condition obtained without consideration of any interface), 
so that bifurcation is always excluded when both conditions 
$T_{nn} \geq 0$ and $T_{tt} \geq 0$ hold true. Using the correct parameter $\alpha=1$, the exclusion condition evidences a term pertaining  to the interface, which allows the bifurcation to occur for both positive $T_{nn}$ and $T_{tt}$.

\subsubsection{Elastic layer on an elastic half-space, in contact through a planar sliding interface} 
\label{sec:layer}

\begin{figure}[!htb]
\centering
\includegraphics[width=16cm]{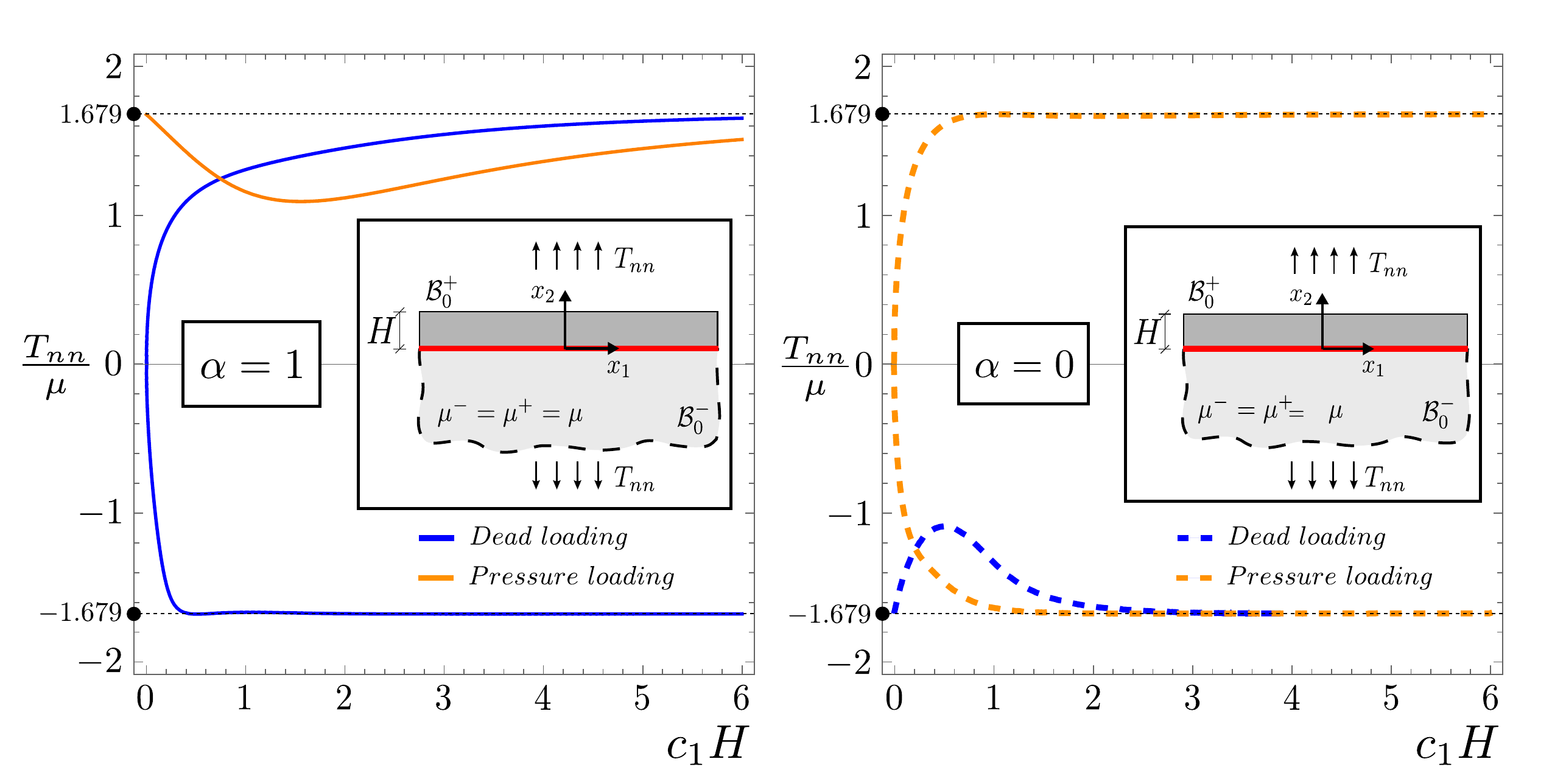}
\caption{\footnotesize Bifurcation of a layer connected to an elastic incompressible half-space through a sliding interface. Both layer and half-space are modelled with the same neo-Hookean material and subject to the same prestress orthogonal to the interface. Both dead and pressure loadings are considered for the two interfacial conditions $\alpha=1$ and $\alpha=0$ (the latter condition is incorrect and included only for comparison).  The normalized bifurcation stress $T_{nn}/\mu$ is reported versus the normalized wavenumber of the bifurcated field $c_{1}H$. Note that for dead load bifurcation in tension is possible only when the correct interfacial condition, $\alpha=1$, is considered.}
\label{fig010}
\end{figure}

An elastic layer (of current thickness $ H $) is considered, connected to an elastic half-space through a planar sliding interface, see the inset in Fig.~\ref{fig010}. Both the layer and the half-space are assumed to obey the same neo-Hookean material model. The system is subject to a uniform biaxial Cauchy prestress state with principal components $ T_{tt} $ and $ T_{nn} $. A reference system $x_1$--$x_2$ is introduced aligned parallel respectively to the unit tangent $\bt$ and normal $\bn$ to the interface.

In addition to the incremental boundary conditions given by Eqs.~\eqref{ball}--\eqref{ballblack} at the sliding interface ($x_{2} = 0$), the decaying condition as $x_2 \to -\infty$, plus the condition holding at the free surface ($x_{2} = H$), have to be enforced. The latter condition differs for dead or pressure loading as follows:
 \begin{itemize}
\item  for dead loading,
\begin{equation}
\dot{S}_{nn}^{+} \left( x_{1},H \right) = \dot{S}_{tn}^{+} \left( x_{1},H \right) = 0 ;
\label{deadL}
\end{equation}
\item for pressure loading,
\begin{equation}
\dot{S}_{nn}^{+} \left( x_{1},H \right) = -  T_{nn} L_{nn} \left( x_{1},H \right) , \qquad
\dot{S}_{tn}^{+} \left( x_{1},H \right) = -  T_{nn} L_{nt} \left( x_{1},H \right).
\label{pressureL}
\end{equation}
\end{itemize}
Imposing the above conditions, a linear homogeneous system is obtained for the bifurcation stress $T_{nn} / \mu$, 
when the longitudinal prestress is assumed null ($T_{tt} / \mu = 0$). The bifurcation stress is reported in Fig.~\ref{fig010} as a function of the wavenumber of the bifurcated field, for both situations of dead loading and pressure loading and for both 
correct and incorrect conditions, respectively, $\alpha=1$ and $\alpha = 0$. 

For pressure loading, a tensile bifurcation is observed, which occurs for both the correct ($\alpha=1$, left in the figure) and incorrect ($\alpha=0$, right in the figure) conditions at the 
interface. A tensile bifurcation for dead loading is possible only when the correct condition $\alpha=1$ is employed, while in the other case the Hill's type 
condition (see Appendix \ref{appendixB}) excludes bifurcations for tensile $T_{nn}$ and null $T_{tt}$. 
In any case, results are strongly different for the correct and incorrect models of interface, showing once again the importance of a correct modelling of interfacial conditions. 

\subsubsection{Two elastic layers} 

\begin{figure}[!htb]
\centering
\includegraphics[width=16cm]{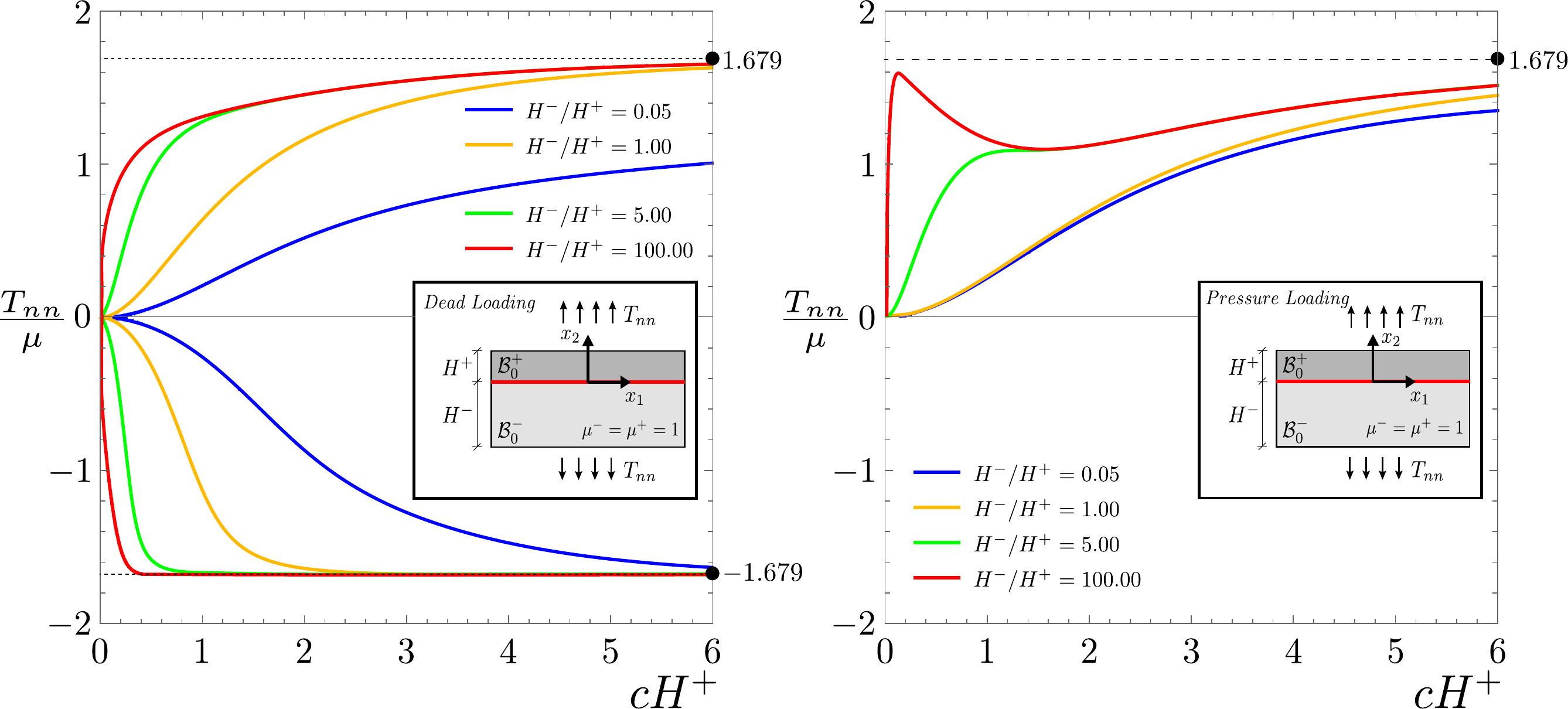}
\caption{\footnotesize Bifurcation of two elastic incompressible layers in contact through a sliding interface. 
Both layers are modelled with the same neo-Hookean material and subject to the same prestress orthogonal to the interface. 
The normalized bifurcation stress $T_{nn}/\mu$ is reported versus the normalized wavenumber of the bifurcated field $c_{1}H^+$, for different values of the thickness ratio $ H^{-}/H^{+}$.}
\label{fig02LAYER}
\end{figure}

Two layers (one denoted by \lq $+$' and the other by \lq $-$'), connected through a planar sliding interface are considered, subject to transverse and longitudinal prestresses $T_{nn}$ and $T_{tt}$.
The transverse stress is assumed to be generated by either a dead, Eqs.~(\ref{deadL}), or a pressure, Eqs.~(\ref{pressureL}), loading (see the insets in Fig.~\ref{fig02LAYER}).  Now only the correct condition $\alpha=1$ is considered, as for $\alpha=0$ the Hill's type condition excludes bifurcation for positive dead loading $T_{nn}$ and null transversal loading, see Appendix \ref{appendixB}.

As in the case of a layer on a half-space ($H^-/H^+\to\infty$), see Section~\ref{sec:layer}, compressive pressure loading, $T_{nn}<0$, does not lead to buckling, 
and tensile dead loading yields a bifurcation.
The results for $H^-/H^+<1$ are included in Fig.~\ref{fig02LAYER} for illustration purposes only, as they correspond to the respective results for the reciprocal value of $H^-/H^+>1$ upon adequate rescaling of $c H^+$.

\section{Bifurcations in Complex Problems Involving a Sliding Interface}
\label{sec04}

A special feature characterizing the presence of sliding interfaces is the appearance of tensile bifurcations, often excluded for other models of interfaces (for instance in the 
perfectly bonded case). These bifurcations are usually hard to be obtained analytically (the simple cases reported in the previous section are of course exceptions), so that the aim of this 
section is to use a finite-element method combined with a linear perturbation analysis to analyze tensile bifurcations occurring under plane strain conditions in a system of two elastic slender blocks 
and a hollow cylinder with an internal coating, in both cases jointed through a sliding interface. 
The former mechanical system is related to the problem of buckling in tension of two elastic rods (Zaccaria et al.,\ 2011), while the latter is related to a  problem of coating detachment.

\subsection{Finite-element treatment}

A mixed formulation is adopted in order to implement incompressible hyperelasticity in plane-strain conditions. 
Quadrilateral 8-node elements are used with quadratic (serendipity) interpolation of displacements and continuous bilinear interpolation of the pressure field that plays the role of a Lagrange multiplier enforcing the incompressibility constraint using the augmented Lagrangian method. 
Standard $3{\times}3$ Gaussian quadrature is applied. 
As in the analytical examples studied in the previous section, the constitutive response is modelled using the incompressible neo-Hookean model.

The sliding interface is modeled as a frictionless bilateral interface in the geometrically-exact finite-deformation setting. 
Quadratic interface elements are used for that purpose with each surface represented by three nodes, so that curved interfaces can be correctly represented. 
The closest-point projection is used to determine the points that are in contact, and the augmented Lagrangian method is used to enforce the bilateral (equality) constraint. 
Those aspects follow the standard concepts used in computational contact mechanics (Wriggers, 2006), except that here bilateral rather than unilateral contact is considered. 
The present implementation employing interface elements is suitable for relatively small, but finite relative sliding. 
This is sufficient for the purpose of bifurcation analysis that is carried out below.

The bifurcation analysis is performed using a linear perturbation technique. 
Specifically, a linear perturbation is applied in the deformed (prestressed) base state that corresponds to a gradually increasing load, and the bifurcation point is detected when the perturbation grows to infinity.

Implementation and computations have been performed using the \emph{AceGen/AceFEM} system
(Korelc, 2009). 
As a verification of the computational scheme, the problem of two elastic half-spaces (Section~\ref{sec:half-spaces}) and the problem of a layer on an elastic half-space (Section~\ref{sec:layer}) have been analyzed, and a perfect agreement with the corresponding analytical solutions has been obtained.

\subsection{Tensile bifurcation of two elastic slender blocks connected through a sliding interface}

\begin{figure}[!htb]
\centering
\includegraphics[width=11cm]{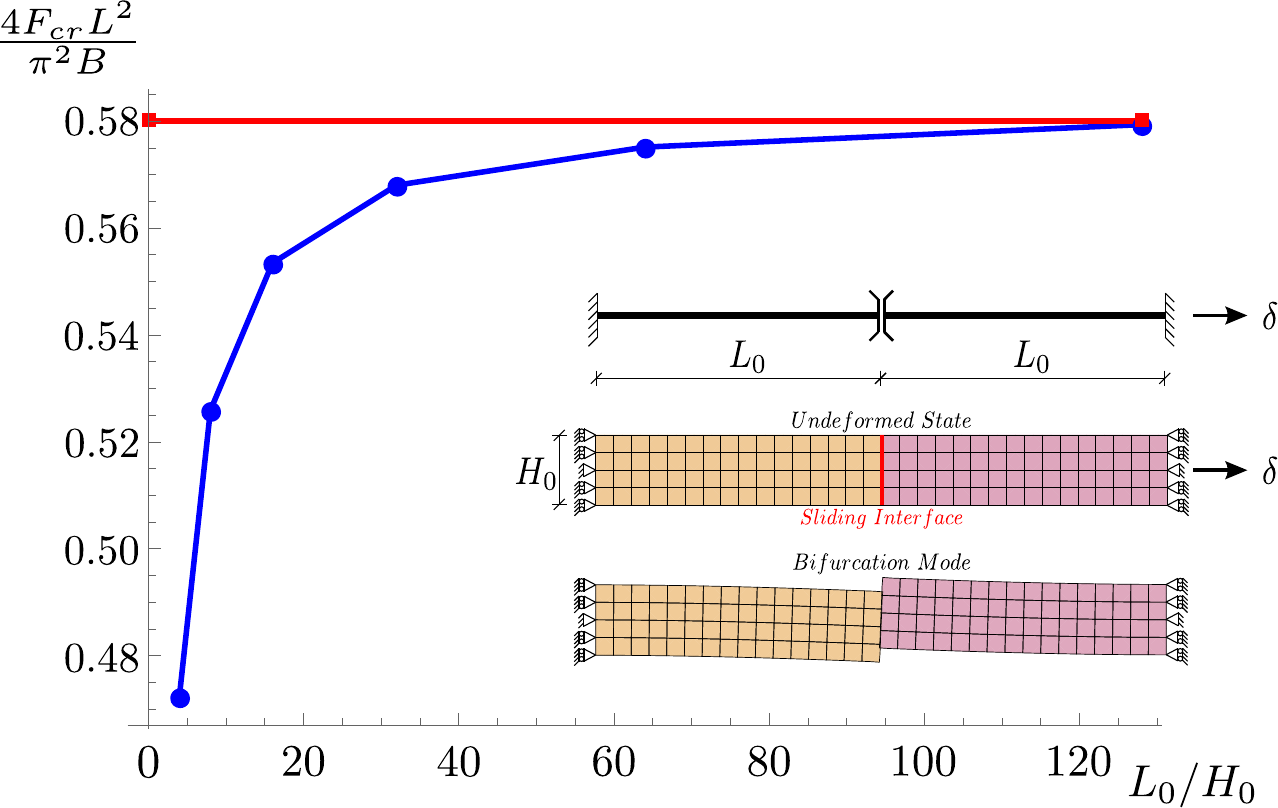}
\caption{\footnotesize{Two identical neo-Hookean rectangular blocks uniformly deformed in tension, jointed through a sliding interface. The blocks have initial length $L_0$, width $H_0$, and shear modulus $ \mu_0 = \mu^{+}_0 = \mu^{-}_0$.
The bifurcation force $F_{cr}$ is made dimensionless through multiplication by the square of the current length $L$ of the blocks and division by the bending stiffness $B$ (per unit thickness) of the blocks calculated with reference to 
their current width $L$. Note that the bifurcation force tends, at increasing length of the block, to the value calculated for two elastic rods in tension of shear stiffness $\mu_0$ (reported with a straight red line). 
}}
\label{fig5}
\end{figure}

As the first numerical example, bifurcation in tension is studied for the problem of two identical elastic rectangular blocks jointed through a frictionless bilateral-contact interface, see the inset in Fig.~\ref{fig5}. 	
The axial displacements are constrained at one support and uniform axial displacement is prescribed at the other support. 
Additionally, in each block, the lateral displacement is constrained at one point in the middle of the support. 
In the base state, the rods are thus uniformly stretched, while the bifurcation mode in tension involves bending of both blocks 
accompanied by relative sliding at the interface, as shown in the inset of Fig.~\ref{fig5}, where the problem scheme, together with the undeformed mesh and the deformed mesh at buckling are reported (the mesh used in the actual computations was finer than that shown in Fig.~\ref{fig5} as an illustration). 

The present problem is, in fact, a continuum counterpart of the problem, studied by Zaccaria et al.\ (2011), of tensile bifurcation of two inextensible elastic Euler--Bernoulli beams clamped at one end and jointed through a slider. 
For that problem, the normalized critical tension force $F_{\it cr}$ has been found equal to $4F_{\it cr}L^2/(\pi^2 B)=0.58$, where $L$ denotes the beam length and $B$ the bending stiffness.

Figure~\ref{fig5} shows the normalized critical force as a function of the initial length-to-height ratio, $L_0/H_0$. 
For consistency, the force has been normalized using the current length $L=\lambda L_0$ and the bending stiffness $B=\mu H^3/3$ (per unit thickness) has been determined in terms of the current height $H=\lambda^{-1}H_0$ and current incremental shear modulus $\mu=\mu_0(\lambda^2+\lambda^{-2})/2$, even though the critical stretch $\lambda$ is close to unity (e.g., $\lambda=1.006$ for $L_0/H_0=4$ and  $\lambda=1.002$ for $L_0/H_0=8$).
The result in Fig.~\ref{fig5} shows that for slender blocks the critical force agrees well with the model of Zaccaria et al.\ (2011), which critical load is reported with a red straight line. 
For thick blocks, the two models differ, for instance, by $20\%$ at $L_0/H_0=4$.

\subsection{Hollow cylinder with internal coating} 

\begin{figure}[!htb]
\centering
\includegraphics[width=15cm]{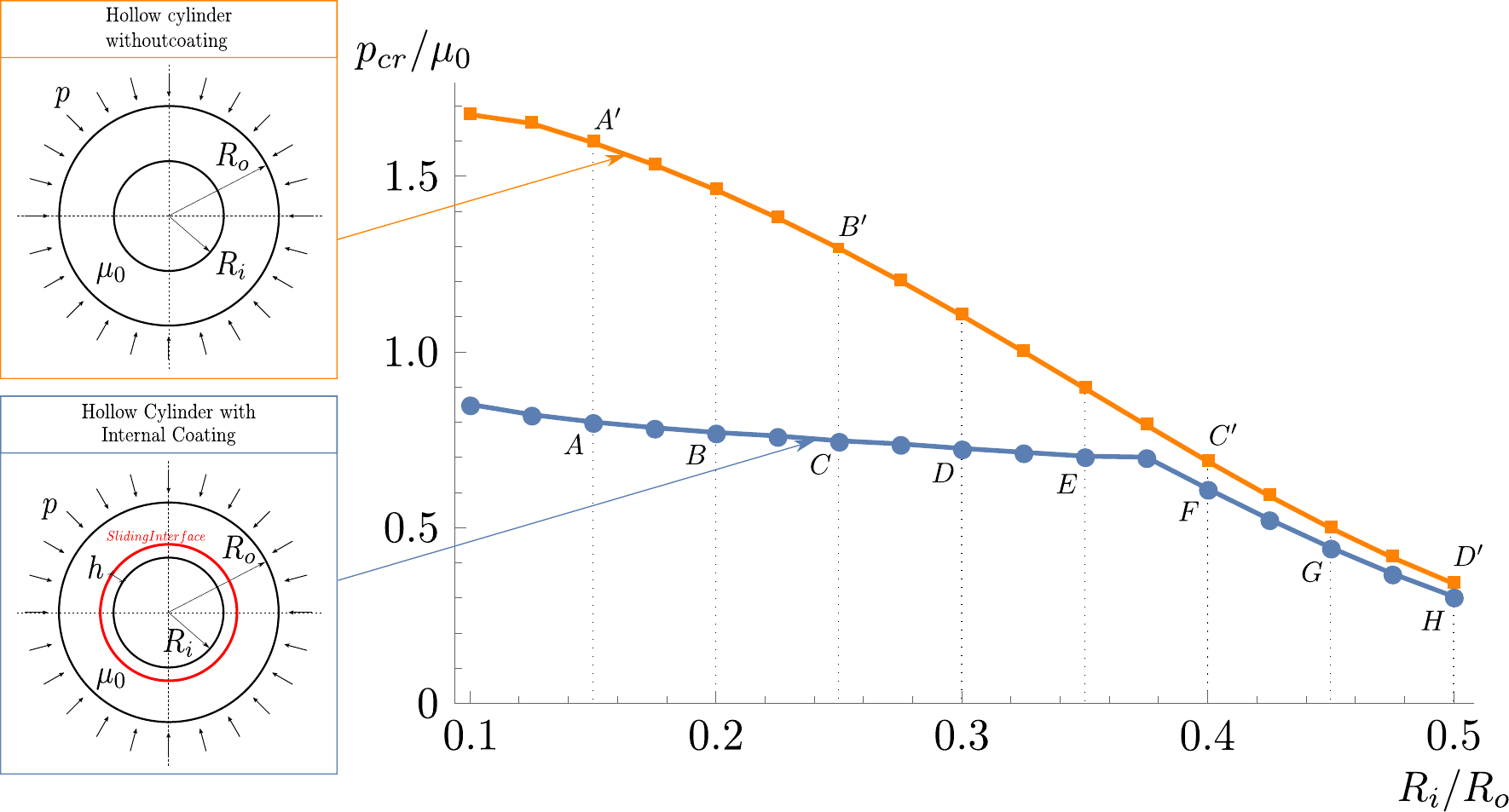}
\caption{\footnotesize{Bifurcation pressure $p_{cr}$, made dimensionless through division by the shear modulus $\mu_{0}$, for a cylinder with (blue line) and without (orange line) internal coating, as a function of the ratio between the inner and outer radii  of the cylinder, $R_i/R_o$. The coating is connected to the cylinder with a sliding interface. Note the strong decrease of the bifurcation pressure due to the presence of the coating.  
}}
\label{crloadcylinder}
\end{figure}

A hollow cylinder is now considered with an internal coating and loaded by a uniform external pressure. 
The cylinder and the coating interact through a frictionless contact interface. 
The geometry is specified by the outer radius $R_o$, the inner radius $R_i$, and the coating thickness $h$ that has been assumed equal to $h=0.01R_o$, see the inset in Fig.~\ref{crloadcylinder}. 
The shear moduli of the tube and coating are equal. The case where the coating is absent is also investigated for comparison. 

\begin{figure}[!htb]
\centering
 \includegraphics[width=14cm]{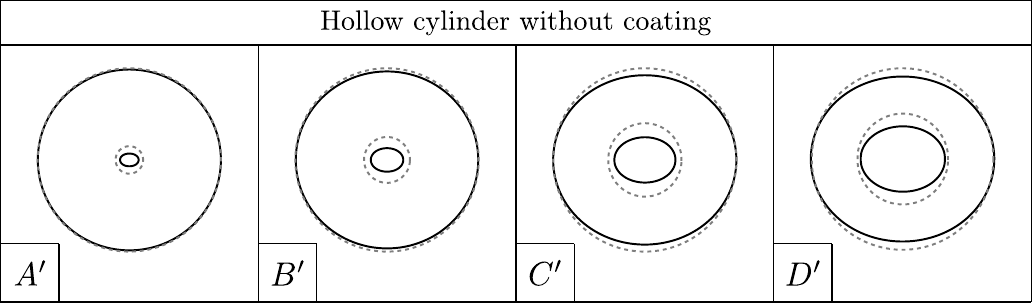}
\caption{\footnotesize{Bifurcation modes for a hollow cylinder (without coating) subjected to an external pressure (dashed lines denote the undeformed configuration, solid lines denote the bifurcation mode in the deformed configuration). 
The bifurcation modes correspond to the loads indicated in Fig.~\ref{crloadcylinder}, to which the letters are referred.
}}
\label{MODEmonolit}
\end{figure}

\begin{figure}[!htb]
\centering
\includegraphics[width=14cm]{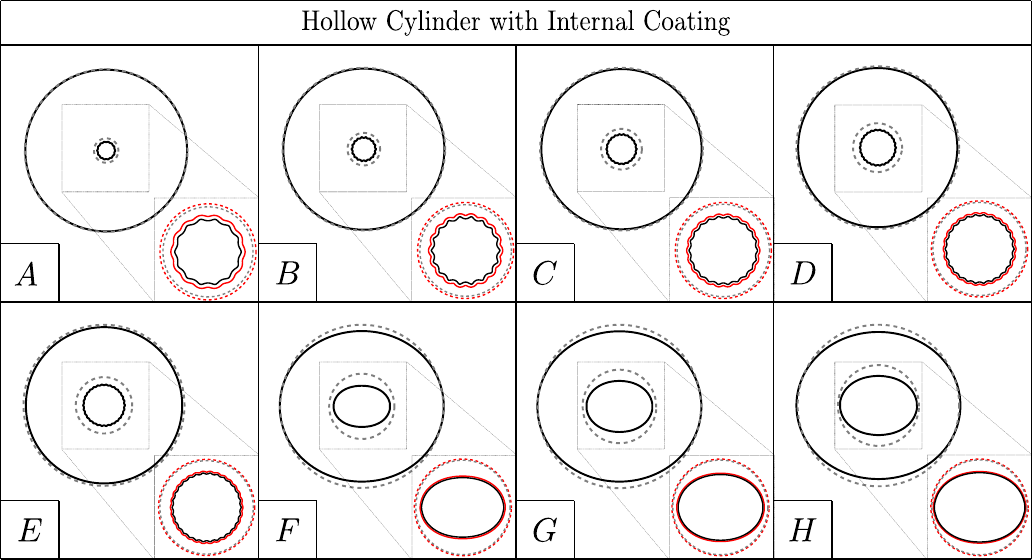}
\caption{\footnotesize{Bifurcation modes for a hollow cylinder with an internal coating jointed through a sliding interface. The cylinder is subjected to an external pressure. Bifurcation modes correspond to the loads indicated in Fig.~\ref{crloadcylinder}, to which the letters are referred. Note that an enlarged detail of the inner, coated surface is reported for each geometry (dashed lines denote the undeformed configuration, solid lines denote the bifurcation mode in the deformed configuration, the sliding interface is denoted in red).
}}
\label{MODEcylinderlayer}
\end{figure}

Figure~\ref{crloadcylinder} shows the critical pressure $p_{\it cr}$ normalized through division by the shear modulus $\mu_0$ as a function of the inner-to-outer radius ratio,  $R_i/R_o$. 
As a reference, the critical load of a hollow cylinder without coating is also included. 
The bifurcation modes are reported in Fig.~\ref{MODEmonolit} for the uncoated and in Fig. \ref{MODEcylinderlayer} for the coated case. 
In the case of coating, two buckling modes are observed depending on the wall thickness. 
For $R_i/R_o$ greater than approximately 0.38, a global buckling mode occurs, as illustrated in Fig.~\ref{MODEcylinderlayer}.
This mode is also characteristic for the uncoated hollow cylinder in the whole range of $R_i/R_o$. 
For the same ratio of $R_i/R_o$ and the same load $p/\mu_0$, the base state is identical for the cylinder with coating and for the uncoated one. 
However, the critical load is different, and, in the global-mode regime, the sliding interface reduces the critical load by approximately 11\%.

A local bifurcation mode is observed for the coated hollow cylinder when $R_i/R_o$ is less than approximately 0.38, as illustrated in Fig.~\ref{MODEcylinderlayer}. 
In this buckling mode, the layer and the inner part of the tube deform in a wave-like fashion, while the outer part of the tube remains intact. 
This mode is thus
similar to the buckling mode characteristic for the layer resting on an elastic half-space, see Section~\ref{sec:layer}, with the difference that here the substrate is curved. 
In the local-mode regime, the critical load is significantly reduced with respect to the uncoated cylinder (which buckles in the global mode). 
For instance, for $R_i/R_o=0.1$, the critical load is reduced by 50\%. 


As a conclusion, the presence of a coating connected with a sliding interface is detrimental to the stability of the system, so that the coating tends to slide and the bifurcation load is strongly lower than that calculated in the case when the coating is absent.

\section{Experimental Evidence of Tensile Bifurcation and Sliding Between Two Soft Solids in Contact Through a Sliding Interface}
\label{sec05}

\begin{figure}[!htb]
\centering
\includegraphics[width=10cm]{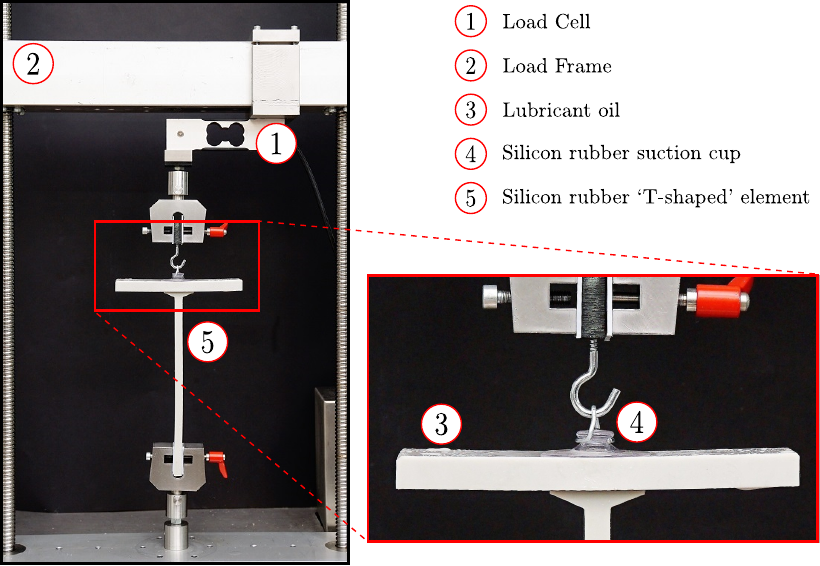}
\caption{\footnotesize{The set-up of an experiment showing a tensile bifurcation involving two soft solids connected trough a sliding interface. A vertical displacement (rotations are left free) is imposed to the head of a suction cup connected to a \lq T-shaped' silicon rubber element. 
A lubricant oil is applied, so that the suction cup can slide along the upper edge of the \lq T' element.
}}
\label{figTestSetUp}
\end{figure}

As mentioned in the introduction, experiments have been designed and realized (in the \lq Instabilities Lab' of the University of Trento), showing a tensile bifurcation which involves 
two soft solids connected through a sliding interface, Fig.~\ref{figTestSetUp}. 

In particular, a \lq T-shaped' silicon rubber element has been manufactured with a \lq stem' having rectangular cross section 
10 mm $\times$ 30 mm (RBSM from Misumi, with 7.4 MPa ultimate tensile strength) and an upper end of dimensions 160 mm $\times$ 10 mm $\times$ 40 mm. 
Three different lengths of the stem have been tested, namely, $L_{1}=210$ mm, $L_{2}=180$ mm, and $L_{3}=150$ mm. 
The upper flat part of the \lq T' has been attached (through a lubricant oil, Omala S4WS 460) to a silicon rubber suction cup. The suction cup has been pulled in tension (by imposing a vertical displacement at a velocity of 0.7 mm/s, with a uniaxial testing machine, Messphysik midi 10). The load and displacement have been measured respectively with a load cell (a MT1041, RC 20kg, from Metler Toledo) and the potentiometric transducer inside the testing machine. 
Data have been acquired with a system NI CompactDAQ, interfaced with Labview (National Instruments). 

The oil used at the suction cup contact allows the suction cup to slide along the upper part of the \lq T' element. Therefore, when the suction cup is pulled, the system initially remains straight and the stem deforms axially. However, at a sufficiently high load, a critical condition is reached and the system buckles. Consequently, the stem of the \lq T' element bends and the suction cup slides along its upper flat end, see Fig.~\ref{figFRAME}. 

This is a simple experiment showing a tensile bifurcation of two soft elastic materials (the \lq T' element and the suction cup), when they are connected through a sliding interface, a phenomenon which is predicted by the model developed in the present paper, in particular by the use of the correct interface conditions (\ref{ball})--(\ref{ballblack}).

Note, however, that the oil does not allow a completely free sliding of the suction cup, so that an initial relative movement at the suction cup--rubber element interface requires the attainment of an initial force, which suddenly decreases when the relative displacement increases and eventually becomes negligible, thus realizing the sliding interfacial conditions analyzed in the present paper. 
This is evident in the load-displacement curves, shown in Fig.~\ref{figCURVES}, two for each tested length. The curves are marked blue for $L=210$ mm, green for $L=180$ mm and red for $L=150$ mm. 
The curves show a peak in the force, followed by steep softening and the final attainment of a steady sliding state, where the junction behaves as a sliding interface. 
The peak forces exhibit a significant scatter which is related to the transition from sticking friction, through mixed lubrication at the onset of sliding, to hydrodynamic lubrication during developed sliding, the latter exhibiting much smaller scatter.

The interest in the developed soft system is that it allows the realization of an element buckling in tension, which is essentially similar to the structural system designed by Zaccaria et al. (2011), but now obtained without the use of rollers or other mechanical devices. 

\subsection{Finite element simulations}

Two-dimensional plane-stress finite element simulations have been performed with Abaqus to validate the model of a sliding interface between two soft materials against the experimental results presented in the previous section.

\begin{figure}[!htb]
\centering
\includegraphics[width=13cm]{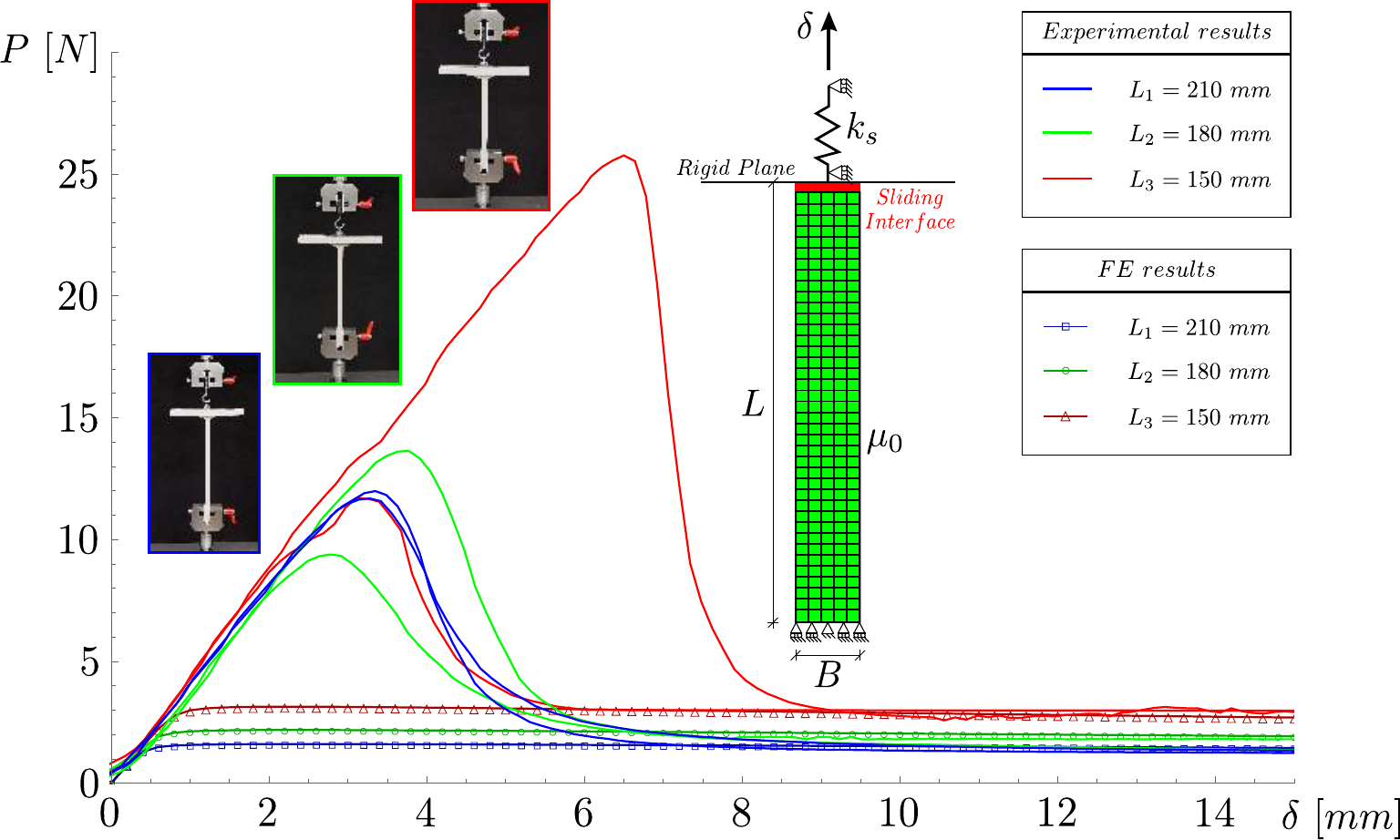}
\caption{\footnotesize{Experimental and simulated load--displacement curves of the structure sketched in the inset for three different lengths of the vertical stem, $L_{1}=210$ mm (red lines), $L_{2}=180$ mm (green lines) and $L_{3}=150$ mm (blue lines). The model of sliding interface correctly captures the post-critical behaviour, where the lubricated contact realized a low friction sliding condition.}
}
\label{figCURVES}
\end{figure}

The geometry is shown in the inset of Fig.~\ref{figCURVES} and consists in a rectangular block of edges $B=10$ mm and $L=\lbrace210,180,150\rbrace$ mm.
The lower edge of the elastic block is clamped, whereas the upper edge is in contact with a rigid plane which can freely rotate and is connected to an elastic spring which models the 
stiffness of the suction cup. 
Contact conditions at the interface between the elastic block and the rigid plane (shown as a red line in the inset of Fig.~\ref{figCURVES}) are prescribed such that a bilateral and frictionless interaction is realized.
An initial imperfection has been introduced, that consists in a rotation of the rigid plane by an angle of $0.5^\circ$. 
The rigid plane is modelled using a two-dimensional 2-node rigid element (R2D2), while the rectangular block is modelled using 4-node bilinear elements with reduced integration and hourglass control (CPS4R element in Abaqus). The material of the elastic block is a neo-Hookean hyperelastic material characterized by a shear modulus $\mu_{0}=7$ MPa. The spring describing the suction cup is a linear elastic spring with stiffness $k_{s}=4.25$ MPa. Displacement boundary conditions (vertical displacement $\delta=15$ mm) are prescribed at the upper end of the elastic spring.

The results of the finite element simulations are shown in Fig.~\ref{figCURVES} as solid lines with markers. It is shown that the finite element model is able to predict correctly the post-critical behaviour. The peak load is not predicted by the model because the effects of the lubricant at the interface (which produces an increase of the load before buckling) are not taken into account.

\section{Conclusions}

A model of sliding interface has been developed for soft solids in sliding contact, a problem of interest in various technologies, exemplified through the design and experimentation on a soft device, which realizes a compliant slider. 
The derived incremental equations are not trivial and differ from previously (and erroneously) employed interface conditions. 
A fundamental simplifying assumption in the model is the bilaterality of the contact, which on the other hand is the key to obtain analytical solutions for several bifurcation problems. 
Some of these solutions have been obtained, which show that: (i.) the interface plays a strong role in the definition of critical conditions, (ii.) 
the interface promotes tensile bifurcations, one of which has been experimentally verified, which cannot be detected if previously used (and erroneous) interfacial conditions are used.

\paragraph{Data accessibility.} This article has no additional data.

\paragraph{Competing interests.} We declare we have no competing interests.

\paragraph{Authors' contributions.} All authors contributed equally to this work and gave final approval for publication.

\paragraph{Acknowledgements.} We thank D. Misseroni for the help with experiments. 

\paragraph{Funding.} A.P., S.S., N.B. gratefully acknowledge financial support from the ERC Advanced Grant \lq Instabilities and nonlocal multiscale modelling of materials'
ERC-2013-ADG-340561-INSTABILITIES. D.B. thanks financial support from the PRIN 2015 ``Multi-scale mechanical models for the design and optimization of micro-structured smart materials and metamaterials'' 2015LYYXA8-006.

\appendix
\renewcommand{\theequation}{\thesection.\arabic{equation}}

\section{An exclusion condition for bifurcation of two solids in contact with a sliding interface}
\setcounter{equation}{0}
\label{appendixB}

\begin{figure}[!htb]
\centering
\includegraphics[width=11cm]{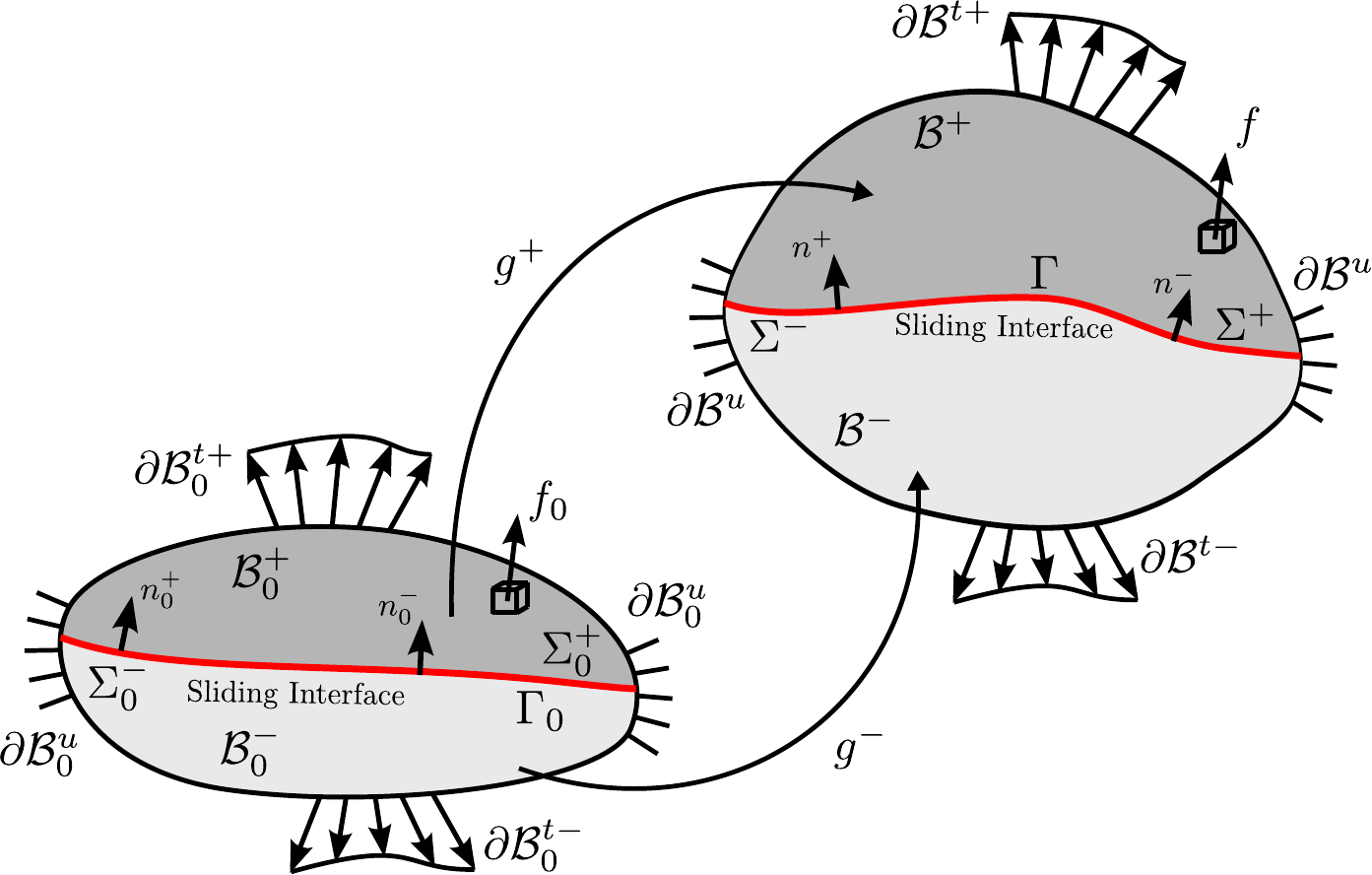}
\caption{\footnotesize{Deformation of a solid containing a sliding interface. $\mB_0$ and $\mB$ denote the reference and current configuration, respectively. 
}}
\label{fig4}
\end{figure}

Following the Hill (1957) generalization of the Kirchhoff proof of uniqueness of the linear theory of elasticity, two incremental solutions are postulated, for the problem sketched 
in Fig.\ \ref{fig4}, 
$\dot{\bx}_\alpha^ {\pm } $, $\dot{\bS}_\alpha ^ {\pm } $ (with $\alpha=1,2$), so that the difference fields $\Delta \dot{\bx}^ {\pm }$, $\Delta \dot{\bS} ^ {\pm } $ are in equilibrium with homogeneous boundary conditions and null body forces. 

Integration of the equilibrium equations for both bodies yields
\begin{equation}
\int_{\mathcal B_0^{\pm}} \left( \Div \Delta \dot{\bS}^{\pm} \right) \scalp \Delta \dot{\bx}^{\pm} = 
\int_{\mathcal B_0^{\pm}} \Div \left( \Delta \dot{\bS}^{\pm T} \Delta \dot{\bx}^ {\pm} \right) - 
\int_{\mathcal B_0^{\pm}} \Delta \dot{\bS}^{\pm} \scalp \Delta \dot{\bF}^ {\pm}
= 0 ,
\end{equation}
so that the divergence theorem provides
\begin{equation}
\int_{\mathcal B_0^{\pm}} \Delta \dot{\bS}^{\pm} \scalp \Delta \dot{\bF}^{\pm} = \mp \int_{\Sigma^{\pm}_{0}} \Delta \dot{\bx}^{\pm} \scalp \Delta \dot{\bS}^{\pm} \bn_{0}.
\label{eint1}
\end{equation}
A sum of the two Eqs.\ \eqref{eint1} yields the following Hill's type exclusion condition for bifurcation
\begin{equation}
\lb{peppa}
\int_{\mathcal B_0} \Delta \dot{\bS} \scalp \Delta \dot{\bF} > - \int_{\Sigma_0} \left( \Delta \dot{\bx}^ {+} \scalp \Delta \dot{\bS} ^ {+} \bn_{0} - \Delta \dot{\bx}^{-} \scalp \Delta \dot{\bS}^{-} \bn_{0} \right) \qquad \forall \Delta \dot{\bS}^\pm, \Delta \dot{\bx}^\pm.
\end{equation}

Before proceeding with the assumptions employed in the present article, the exclusion condition (\ref{peppa}) is specialized to the case of the 
\lq spring-type' interface introduced by Suo et al.\ (1992) and employed also by Bigoni et al.\ (1997). 
This interface is charaterized by: (i.) full continuity of the nominal incremental tractions across the interface and (ii.) a linear interfacial constitutive law of the type
\beq
\lb{stronz}
\dot{\bS}^-\bn_0 = \bH \jump{ \dot{\bx}},
\eeq
where $\bH$ is a constitutive tensor [note that in the notation of the present paper there is a sign differing in equation (\ref{stronz}) from Suo et al.\ (1992)]. Using the two above conditions (i.) and (ii.) in equation (\ref{peppa}), the exclusion condition becomes
\begin{equation}
\lb{stronz2}
\int_{\mathcal B_0} \Delta \dot{\bS} \scalp \Delta \dot{\bF} +
\int_{\Sigma_0} \jump{ \Delta\dot{\bx}} \scalp \bH \jump{ \Delta\dot{\bx}} > 0 \qquad \forall \Delta \dot{\bS}, \Delta \dot{\bx}.
\end{equation}
Equation (\ref{stronz2}) shows that for a positive-definite interfacial tensor $\bH$ (in other words excluding softening interfaces) the term pertaining to the interface is always positive. It may be easily concluded that:
\begin{quote}
When the incremental constitutive response of a solid is governed by a positive definite tensor (as for instance for a Mooney--Rivlin material subject to non-negative principal stresses) bifurcation is always excluded for 
mixed boundary conditions of dead loading and imposed displacements 
{\it even in the presence of positive-definite interfaces of the type introduced by Suo et al.\ (1992)}.
\end{quote}
For instance, in a case in which all principal stresses are positive or null (as it happens in a tensile problem of the type experimentally investigated in this paper) bifurcations are excluded. 
In order to substantiate the above statement with an example, consider two elastic blocks made up of Mooney--Rivlin material connected through a planar interface of the type proposed by Suo et al.\ (1992) without softening. 
If these blocks will be pulled in tension with a dead loading, the condition (\ref{stronz2}) excludes all possible bifurcations. But the bifurcation will occur in reality, as the T-problem shows. This bifurcation is found if the interface is replaced with a sliding interface of the type described by equations (\ref{ball})--(\ref{ballblack}).

The following assumptions are now introduced:
\begin{itemize}
\item a Lagrangean formulation is assumed with the current  state taken as reference, so that $\mB_0 \equiv \mB$ and $\Sigma_0 \equiv \Sigma$; 
\item plane strain deformation in the plane $x_1$--$x_2$ prevails; 
\item a planar interface is assumed, so that $\bn_{0}=\bn$ and $\bt_{0}=\bt$; 
\item the material is prestressed by a uniform Cauchy stress with principal components $T_{tt}$ and $T_{nn}$;
\item the constitutive equation of the material is incrementally linear
\begin{equation}
\dot{\bS} = {\bmE}[\dot{\bF}] \quad \text{for compressible material},
\end{equation}
\begin{equation}
\dot{\bS} = {\bmE}[\dot{\bF}] + \dot{p} \Id  \quad \text{for incompressible material}.
\end{equation}
\end{itemize} 
Then Eq.~(\ref{peppa}) becomes
\begin{equation}
\lb{ocio}
\int_{\mathcal B} \Delta \dot{\bS} \scalp \Delta \bL >  
- \int_{\Sigma} \left( \Delta v^{+}_t \Delta \dot{S}^{+}_{tn} + \Delta v^{+}_n \Delta \dot{S}^{+}_{nn} -  \Delta v^{-}_t \Delta \dot{S}^{-}_{tn} - \Delta v^{-}_n \Delta \dot{S}^{-}_{nn}  \right), 
\end{equation}
where $\bv$ is the incremental displacement and $\bL$ its gradient and repeated indices are not summed.

Introducing the fourth-order elastic tensor ${\bmE}$ and using Eqs.\ (\ref{ball}) and (\ref{ballred}), Eq.~(\ref{ocio}) can be rewritten as 
\begin{equation}
\int_{\mathcal B} \Delta \bL \scalp {\bmE} [ \Delta \bL ] >  
- \int_{\Sigma} \left( \left( \Delta v^{+}_t - \Delta v^{-}_t \right)  \Delta \dot{S}_{tn} + \Delta v_n \left( \Delta \dot{S}^{+}_{nn} - \Delta \dot{S}^{-}_{nn} \right)  \right).
\end{equation}
Finally, using Eqs.~(\ref{ballgreen}) and (\ref{ballblack}), the condition for excluding bifurcation in an elastic solid containing a sliding interface becomes
\begin{equation}
\int_{\mathcal B} \grad \bv \scalp {\bmE} [\grad \bv] - \alpha T_{nn} \int_{\Sigma} \left( v_n \jump{v_{t,t}} - \jump{v_t} v_{n,t} \right) > 0,
\label{exclcond}
\end{equation}
holding for all (not identically zero) continuous and piecewise continuously twice differentiable velocity fields $\bv$  satisfying homogeneous conditions on the part of the boundary where incremental displacements are prescribed and assuming arbitrary values on $\Sigma$, but with the normal component satisfying $v_n^+=v_n^-$. 

The parameter $\alpha$ in Eq.~(\ref{exclcond}) highlights the difference between the correct interface conditions ($\alpha=1$) derived in the present work and the incorrect interface conditions ($\alpha=0$) assumed by Steif (1990).

In the special case in which $T_{nn}=0$, Eq.~(\ref{exclcond}) reduces to the Hill exclusion condition
\begin{equation}
\lb{pallle}
\int_{\mathcal B} \grad \bv \scalp {\bmE} [\grad \bv] > 0,
\end{equation}
showing that for a positive definite incremental elastic tensor ${\bmE}$ the incremental solution is unique, whenever the sliding interface 
is free of normal prestress, otherwise bifurcation is not a-priori excluded. 
When the incorrect assumption $\alpha=0$ is made, condition (\ref{pallle}) is obtained independently of the value of $T_{nn}$, thus excluding bifurcation for positive definite ${\bmE}$. 
Positive definiteness of ${\bmE}$ is equivalent to the requirement that the principal prestresses 
$T_1$, $T_2$, and $T_3$ (which enter in the definition of ${\bmE}$) 
satisfy all the inequalities $T_1+T_2>0$, $T_1+T_3>0$, $T_2+T_3>0$ or, for uniaxial tension $T_1>0$ with $T_2=T_3=0$ (Hill, 1967; Bigoni, 2012).

\end{document}

%% file: definitions.tex
\newcommand{\beq}{\begin{equation}}
\newcommand{\eeq}{\end{equation}}
\newcommand{\lb}{\label}
\newcommand{\beqar}{\begin{eqnarray}}
\newcommand{\eeqar}{\end{eqnarray}}
\newcommand{\bit}{\begin{itemize}}
\newcommand{\eit}{\end{itemize}}
\newcommand{\benum}{\begin{enumerate}}
\newcommand{\eenum}{\end{enumerate}}
\newcommand{\barr}{\begin{array}}
\newcommand{\earr}{\end{array}}

\def\scalp{\mbox{\boldmath $\, \cdot \,$}}

\def\XXint#1#2#3{{\setbox0=\hbox{$#1{#2#3}{\int}$}
   \vcenter{\hbox{$#2#3$}}\kern-.5\wd0}}

\def\b0{\mbox{\boldmath $0$}}

\def\bg{\mbox{\boldmath $g$}}

\def\bn{\mbox{\boldmath $n$}}

\def\bt{\mbox{\boldmath $t$}}
\def\bu{\mbox{\boldmath $u$}}
\def\bv{\mbox{\boldmath $v$}}

\def\bx{\mbox{\boldmath $x$}}

\def\bF{\mbox{\boldmath $F$}}

\def\bH{\mbox{\boldmath $H$}}
\def\bI{\mbox{\boldmath $I$}}

\def\bL{\mbox{\boldmath $L$}}
\def\bM{\mbox{\boldmath $M$}}

\def\bS{\mbox{\boldmath $S$}}
\def\bT{\mbox{\boldmath $T$}}

\def\Id{\mbox{\boldmath $I$}}

\def\f0{\ensuremath{\mathbb{O}}}

\newcommand{\mB}{\ensuremath{\mathcal{B}}}

\newcommand{\bmE}{\mbox{\boldmath $\mathcal{E}$}}

\newcommand{\tr}{\mathop{\mathrm{tr}}}

\newcommand{\Div}{\mathop{\mathrm{Div}}}
\newcommand{\grad}{\mathop{\mathrm{grad}}}

\def\IJSS{{\it Int.\ J.\ Solids Struct.}\ }

%% file: sliding_interface_29-11-2017.bbl
\begin{thebibliography}{}

\bibitem{ATESH} Ateshian, G.A., 2009. The role of interstitial fluid pressurization in articular cartilage lubrication.  J. Biomech. 42, 1163–1176.

\bibitem{BIGBOOK} Bigoni, D., 2012. Nonlinear Solid Mechanics. Cambridge University Press, New York.

\bibitem{bigoni2001} Bigoni, D. and Gei, M., 2001. Bifurcations of a coated, elastic cylinder. Int. J. Solids Struct., 38, 5117-5148.

\bibitem{bigoni1997} Bigoni, D., Ortiz, M. and Needleman, A., 1997. Effect of interfacial compliance on bifurcation of a layer bonded to a substrate. J. Mech. Phys. Solids 34, 4305-4326.

\bibitem{ciarletta} Ciarletta, P. and Destrade, M. 2014. Torsion instability of soft solid cylinders. IMA J. Appl. Math. 79, 804-819. 

\bibitem{cristescu} Cristescu, N.D., Craciun,  E.M. and Soos, E., 2004. Mechanics of elastic composites. Boca Raton. Chapman \& Hall/CRC.

\bibitem{deboo} deBotton, G., Bustamante, R. and Dorfmann, A. 2013. Axisymmetric bifurcations of thick spherical shells under inflation and compression. \IJSS 50, 403-413. 

\bibitem{destr} Destrade, M., Fu, Y. and Nobili, A. 2016. Edge wrinkling in elastically supported pre-stressed incompressible isotropic plates. Proc. Royal Soc. A 472, 20160410. 

\bibitem{destrade} Destrade, M. and Merodio, J. 2011. Compression instabilities of tissues with localized strain softening. Int. J. Applied Mechanics 3, 69-83.

\bibitem{deto} De Tommasi, D., Puglisi, G., Saccomandi, G. and Zurlo, G. 2010. Pull-in and wrinkling instabilities of electroactive dielectric actuators. J. Physics D 43, 325501.

\bibitem{Dowaikh} Dowaikh, M.A. and Ogden, R.W., 1991. Interfacial waves and deformations in prestressed elastic media. Proc. Roy. Soc. A 433, 313-328.

\bibitem{DowsonD} Dowson, D., 2012. Bio-tribology. Faraday Discuss. 156, 9-30.

\bibitem{DUNN} Dunn, A.C., Tichy, J.A., Uruena, J.M. and Sawyer, W.G., 2013. Lubrication regimes in contact lens wear during a blink. Tribol. Int. 63, 45-50.

\bibitem{fu2} Fu, Y.B. and Cai, Z.X. 2015. An asymptotic analysis of the period-doubling secondary bifurcation in a film/substrate bilayer. SIAM J. Appl. Math. 75, 2381-2395.

\bibitem{fu} Fu, Y.B. and Ciarletta, P. 2015. Buckling of a coated elastic half-space when the coating and substrate have similar material properties Proc. Royal Soc. A 471, 20140979.

\bibitem{hill1957} Hill, R., 1957. On uniqueness and stability in the theory of finite elastic strain. J. Mech. Phys. Solids 5, 229-241.

\bibitem{hill1967}  Hill, R., 1967. Eigenmodal deformations in elastic/plastic continua. J. Mech. Phys. Solids 15, 371-386.

%

\bibitem{korelc2009} Korelc, J., 2009. Automation of primal and sensitivity analysis of transient coupled problems. Comput. Mech. 44, 631-649.

\bibitem{leroy} Leroy, Y.M. and Triantafyllidis, N., 1996. Stability of a frictional, cohesive layer on a viscous substratum: variational formulation and asymptotic solution. J. Geophys. Res. 101, B8 795-811.

\bibitem{liang} Liang, X. and Cai, S. (2015) Gravity induced crease-to-wrinkle transition in soft materials. Appli. Phys. Letters 106, 041907.


\bibitem{destrade} Ottenio, M., Destrade, M. and Ogden, R.W., 2007. Acoustic waves at the interface of a pre-stressed incomplessible elastic solid and a viscous fluid. Int. J. Non-linear Mech. 42, 310-320.

\bibitem{riccob} Riccobelli, D. and Ciarletta, P. 2017. Rayleigh-Taylor instability in soft elastic layers. Phil. Trans. Royal Soc. A 375, 20160421. 

\bibitem{STEIF3} Steif, P.S., 1990. Interfacial instabilities in an unbonded layered solid. Int. J. Solids Struct. 26, 915-925.

\bibitem{steigm} Steigmann, D.J. and Ogden, R.W. 2014. Classical plate buckling theory as the small-thickness limit of three-dimensional incremental elasticity. ZAMM 94, 7-20. 

\bibitem{SS1} Stupkiewicz, S. and Marciniszyn, A., 2009. Elastohydrodynamic lubrication and finite configuration changes in reciprocating elastomeric seals. Tribol. Int. 42, 615-627.

\bibitem{suo} Suo, Z., Ortiz, M. and Needleman, A., 1992. Stability of solids with interfaces. J. Mech. Phys. Solids 40, 613-640. 

\bibitem{SS2} Temizer, I. and Stupkiewicz, S., 2016. Formulation of the Reynolds equation on a time-dependent lubrication surface. Proc. Roy. Soc. A. 472, 20160032.

\bibitem{tsu} Tsuchida, E., Mura, T. and Dundurs, J., 1986. The elastic field of an elliptic inclusion with a slipping interface. J. Appl. Mech. 53, 103-107.

\bibitem{WRIBOOK} Wriggers, P., 2006. Computational Contact Mechanics. Springer-Verlag, Berlin Heidelberg.

\bibitem{ZACC} Zaccaria, D., Bigoni, D., Noselli, G. and Misseroni, D., 2011. Structures buckling under tensile dead load. Proc. Roy. Soc. A. 467, 1686-1700.

\end{thebibliography}
